\documentclass[12pt,a4paper]{article}

\usepackage{amsmath}
\usepackage{mathtools}
\usepackage{amsfonts}
\usepackage{amssymb}
\usepackage{graphicx}
\usepackage{color}
\usepackage{booktabs}
\usepackage{inputenc}
\usepackage{soul}
\usepackage[T1]{fontenc}
\usepackage{mathrsfs}
\usepackage{enumerate}
\usepackage{url}
\usepackage{cancel,dsfont} 
\usepackage[normalem]{ulem} 
\usepackage[numbers,compress,square]{natbib} 
\usepackage{soul}
\usepackage{multirow}
\usepackage{soul}
\usepackage{cancel}
\usepackage[dvipsnames]{xcolor}

\newcommand{\eea}{\end{eqnarray}}
\newcommand{\bea}{\begin{eqnarray}}
\newcommand{\be}{\begin{equation}}
\newcommand{\ee}{\end{equation}}

\newcommand{\kvec}{\vec{k}}
\newcommand{\qvec}{\vec{q}}

\newcommand{\Phip}{\Phi_{\rm p}}
\newcommand{\fnl}{f_{\rm NL}}

\newcommand{\andd}{\ , \quad \text{and}  \quad}

\newcommand{\eqn}[1]{Eq.~(\ref{#1})}

\newcommand{\secref}[1]{Sec.~\ref{#1}}

\newcommand{\appref}[1]{App.~\ref{#1}}
\newcommand{\figref}[1]{Fig.~\ref{#1}}
\newcommand{\tabref}[1]{Tab.~\ref{#1}}

\newcommand{\xvec}{\vec{x}}
\newcommand{\half}{\frac{1}{2}}

\def\be{\begin{equation}}
\def\ee{\end{equation}}

\def\knl{k_{\rm NL}}

\newcommand{\km}{k_{\rm M}}

\def\hinvMpc{h\,{\rm Mpc}^{-1}}
\def\Mpcinvh{{\rm Mpc}/h}

\newcommand{\kmax}{k_{\rm max }}

\newcommand{\code}[1]{\texttt{#1}}

\usepackage[colorlinks,bookmarks]{hyperref}
\definecolor{linkblue}{rgb}{0,0,0.8}
\definecolor{linkgreen}{rgb}{0,0.5,0}

\hypersetup{pdfpagemode=UseNone, pdfstartview=FitH, linkcolor=linkblue,
            citecolor=linkgreen, urlcolor=linkblue}

\begin{document}

\vspace*{-25mm}

\begin{flushright}
{\small NUHEP-TH/22-02}
\end{flushright}

\begin{center}

{\Large \bf Limits on primordial non-Gaussianities\\[0.3cm] from BOSS galaxy-clustering data}  \\[0.7cm]

{\large   Guido D'Amico${}^{1,2}$,  Matthew Lewandowski${}^{3}$,\\[0.3cm] Leonardo Senatore${}^{4}$, and  Pierre Zhang${}^{5,6,7,4}$ \\[0.7cm]}

\end{center}

\begin{center}

\vspace{.0cm}

\begin{small}

{ { \sl $^{1}$ Department of Mathematical, Physical and Computer Sciences,\\ University of Parma, 43124 Parma, Italy}}
\vspace{.05in}

{ { \sl $^{2}$ INFN Gruppo Collegato di Parma, 43124 Parma, Italy}}
\vspace{.05in}

{ { \sl $^{3}$ Department of Physics and Astronomy,\\ Northwestern University, Evanston, IL 60208}}
\vspace{.05in}

{ { \sl $^{4}$ Institut fur Theoretische Physik, ETH Zurich,
8093 Zurich, Switzerland}}
\vspace{.05in}

{ { \sl $^{5}$ Department of Astronomy, School of Physical Sciences, \\
University of Science and Technology of China, Hefei, Anhui 230026, China}}
\vspace{.05in}

{ { \sl $^{6}$ CAS Key Laboratory for Research in Galaxies and Cosmology, \\
University of Science and Technology of China, Hefei, Anhui 230026, China}}
\vspace{.05in}

{ { \sl $^{7}$ School of Astronomy and Space Science, \\
University of Science and Technology of China, Hefei, Anhui 230026, China}}
\vspace{.05in}

\end{small}

\vspace{.3cm}

\end{center}

%
%
%
%
%
%
%
%
%
%
%
%
%
%
%
%
%

\vspace{.5in}
\hrule \vspace{0.3cm}
{\normalsize  \noindent \textbf{Abstract}
\vspace{.1in}

\noindent We analyze the power spectrum and the bispectrum of BOSS galaxy-clustering data using the prediction from the Effective Field Theory of Large-Scale Structure at  one-loop  order for {\it both} the power spectrum {\it and} the bispectrum. With $\Lambda$CDM parameters fixed to Planck preferred values, we set limits on three templates of non-Gaussianities predicted by many inflationary models: the equilateral, the orthogonal, and the local shapes. After validating our analysis against simulations, we find {$\fnl^{\rm equil.}= 207 \pm 292$},  $\fnl^{\rm orth.}= -68 \pm 73$, $\fnl^{\rm loc.}= 52 \pm 34$, at $68\%$ {confidence level}. These bispectrum-based constraints from Large-Scale Structure, not far from the ones of WMAP,  suggest promising results from upcoming surveys.

\noindent

\vspace{0.3cm}}
\hrule

\vspace{0.3cm}
\newpage

\tableofcontents

\section{Introduction\label{sec:intro}}

The SDSS-III Baryon Oscillation Spectroscopic Survey (BOSS) has provided us with a fantastic mapping of the clustering of galaxies in the nearby Universe~\cite{BOSS:2016wmc}. 
The BOSS data, although modest in volume compared to forthcoming experiments such as DESI~\cite{2013arXiv1308.0847L} or Euclid~\cite{Amendola:2012ys}, are remarkable as they have revealed, and continue revealing, a wealth of cosmological information from the large-scale structure of the Universe.   
\vspace{0.5cm}

In the last few years, a series of works have applied the Effective Field Theory of Large-Scale Structure~(EFTofLSS) prediction at one-loop order to analyze the BOSS galaxy full shape (FS) of the Power Spectrum (PS)~\cite{DAmico:2019fhj,Ivanov:2019pdj,Colas:2019ret}, and, more recently, of the correlation function~\cite{Zhang:2021yna,Chen:2021wdi} {(using the EFT model developed first in \cite{Perko:2016puo})}. 
Ref.~\cite{DAmico:2019fhj}  has also analyzed the BOSS galaxy-clustering bispectrum monopole using the tree-level prediction (see also~\cite{Philcox:2021kcw} for a recent generalization).
By imposing a prior on Big Bang Nucleosynthesis, all $\Lambda$CDM cosmological parameters have been measured from these data. The precision reached on some of these parameters is remarkable. For example, the present amount of matter, $\Omega_m$, and the Hubble constant (see also~\cite{Philcox:2020vvt,DAmico:2020kxu} for subsequent refinements) have error bars that are not far from the ones obtained from the Cosmic Microwave Background (CMB)~\cite{Planck:2018vyg}.
Clustering and smooth quintessence models have also been investigated, finding $\lesssim 5\%$ limits on the dark energy equation of state $w$ parameter using only late-time measurements~\cite{DAmico:2020kxu,DAmico:2020tty}, which is again not far from the ones obtained with the CMB~\cite{Planck:2018vyg}.  The measurements of the Hubble constant provide a new, CMB-independent, method for determining this parameter~\cite{DAmico:2019fhj}, and it is already comparable in precision with the method based on  the cosmic ladder~\cite{Riess:2019cxk,Freedman:2019jwv} and CMB.
Such a tool has allowed us to shed much light on the success (or lack thereof) of some models that were proposed to alleviate the tension in the Hubble measurements (see e.g.~\cite{Verde:2019ivm}) between the CMB and cosmic ladder~\cite{DAmico:2020ods,Ivanov:2020ril} (see also~\cite{Niedermann:2020qbw,Smith:2020rxx}).

These results required an intense and years-long line of study to develop the EFTofLSS from the initial formulation to the level {that allows it} to be applied to data.  {To summarize this situation, we therefore provide the following footnote which gives a compact accounting of the crucial works that paved the way for the current work.}
Even though some of the mentioned papers are not strictly required to analyze the data, we, and we believe probably {anybody} else, would {not} have applied the EFTofLSS to data without all these intermediate results.\footnote{The initial formulation of the EFTofLSS was performed in Eulerian space in~\cite{Baumann:2010tm,Carrasco:2012cv}, and subsequently extended to Lagrangian space in~\cite{Porto:2013qua}.
The dark matter power spectrum has been computed at one-, two- and three-loop orders in~\cite{Carrasco:2012cv, Carrasco:2013sva, Carrasco:2013mua, Carroll:2013oxa, Senatore:2014via, Baldauf:2015zga, Foreman:2015lca, Baldauf:2015aha, Cataneo:2016suz, Lewandowski:2017kes,Konstandin:2019bay}.
These calculations were accompanied by some  theoretical developments of the EFTofLSS, such as a careful understanding of renormalization~\cite{Carrasco:2012cv,Pajer:2013jj,Abolhasani:2015mra} (including rather-subtle aspects such as lattice-running~\cite{Carrasco:2012cv} and a better understanding of the velocity field~\cite{Carrasco:2013sva,Mercolli:2013bsa}), of several ways for extracting the value of the counterterms from simulations~\cite{Carrasco:2012cv,McQuinn:2015tva}, and of the non-locality in time of the EFTofLSS~\cite{Carrasco:2013sva, Carroll:2013oxa,Senatore:2014eva}.
These theoretical explorations also include an enlightening study in 1+1 dimensions~\cite{McQuinn:2015tva}.
An IR-resummation of the long displacement fields had to be performed in order to reproduce the Baryon Acoustic Oscillation (BAO) peak, giving rise to the so-called IR-Resummed EFTofLSS~\cite{Senatore:2014vja,Baldauf:2015xfa,Senatore:2017pbn,Lewandowski:2018ywf,Blas:2016sfa}. 
Accounts of baryonic effects were presented in~\cite{Lewandowski:2014rca,Braganca:2020nhv}. The dark-matter bispectrum has been computed at one-loop in~\cite{Angulo:2014tfa, Baldauf:2014qfa}, the one-loop trispectrum in~\cite{Bertolini:2016bmt}, and the displacement field in~\cite{Baldauf:2015tla}.
The lensing power spectrum has been computed at two loops in~\cite{Foreman:2015uva}.
Biased tracers, such as halos and galaxies, have been studied in the context of the EFTofLSS in~\cite{ Senatore:2014eva, Mirbabayi:2014zca, Angulo:2015eqa, Fujita:2016dne, Perko:2016puo, Nadler:2017qto,Donath:2020abv} (see also~\cite{McDonald:2009dh}), the halo and matter power spectra and bispectra (including all cross correlations) in~\cite{Senatore:2014eva, Angulo:2015eqa}. Redshift space distortions have been developed in~\cite{Senatore:2014vja, Lewandowski:2015ziq,Perko:2016puo}. 
Neutrinos have been included in the EFTofLSS in~\cite{Senatore:2017hyk,deBelsunce:2018xtd}, clustering dark energy in~\cite{Lewandowski:2016yce,Lewandowski:2017kes,Cusin:2017wjg,Bose:2018orj}, and primordial non-Gaussianities in~\cite{Angulo:2015eqa, Assassi:2015jqa, Assassi:2015fma, Bertolini:2015fya, Lewandowski:2015ziq, Bertolini:2016hxg}.
Faster evaluation schemes for the calculation of some of the loop integrals have been developed in~\cite{Simonovic:2017mhp}.
Comparison with high-quality $N$-body simulations to show that the EFTofLSS can accurately recover the cosmological parameters have been performed in~\cite{DAmico:2019fhj,Colas:2019ret,Nishimichi:2020tvu,Chen:2020zjt}}

\begin{figure}[t!]
\hspace{.9in}
\hspace{0.2in}
\includegraphics[width=.55\textwidth]{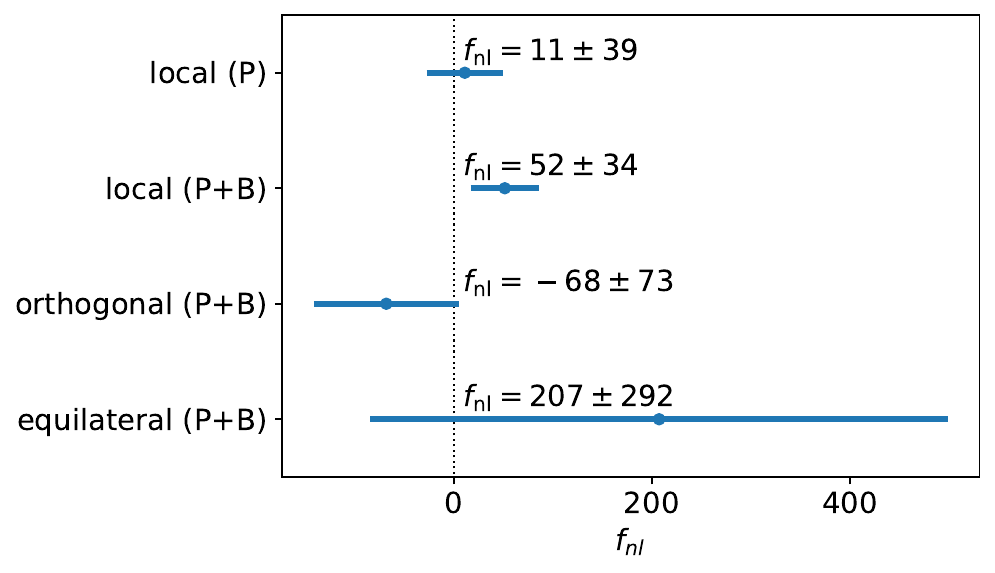}
\caption{\small  Summary plot of the 68\%  confidence level (CL) constraints on primordial non-Gaussianities obtained in this work. $P$ and $B$ represent the fact that the analysis uses either the power spectrum or the power spectrum and bispectrum. We find no evidence of primordial non-Gaussianity.
}
\label{fig:fnl_summary}
\end{figure}

\vspace{0.5cm}

An observable that has been so-far unexplored in galaxy-clustering data analyses from the EFTofLSS is the non-Gaussianity of the primordial fluctuations.  {Non-Gaussianity in the context of the CMB has a long and influential history, and, as we will see below, has thus far provided the most stringent constraints on non-Gaussianity (see for example \cite{WMAP:2003ivt, Creminelli:2005hu, Yadav:2007yy, Senatore:2009gt, Planck:2019kim}).  In the same vein, non-Gaussianity has been explored in the context of LSS mostly through the so-called scale-dependent bias \cite{Slosar:2008hx, Castorina:2019wmr, Mueller:2021tqa}, but also the equilateral shape \cite{Xia:2011hj}, and we compare with many of those important works in \secref{resutltssec}.  }

Some inflationary models predict that the primordial fluctuations have a sizable non-Gaussian distribution. Though there are in principle many potential forms of non-Gaussianities, an extremely large class of models is covered by focusing on the three-point function, and by using three parameterizations for it. They are called `local' (see~\cite{Bernardeau:2002jy} and references therein), `equilateral'~\cite{Creminelli:2005hu}, and `orthogonal'~\cite{Senatore:2009gt}, and they correspond to the following forms of the three-point function of the primordial gravitational field, $\Phip$:\footnote{{Here and elsewhere, $\kvec$ is the wavevector, or Fourier partner of the position $\xvec$.  We also write $k \equiv | \kvec|$ for the magnitude of vectors.  }}
\be\label{eq:3pt}
\langle\Phip(\vec k_1)\Phip(\vec k_2)\Phip(\vec k_3)\rangle=  (2\pi)^3\delta^{(3)}(\vec k_1+\vec k_2+\vec k_3)\,\fnl^{(i)}\, B^{(i)}_{\Phip} (  k_1 , k_2 , k_3 )
\ee
\begin{align}
\begin{split}\label{eq:shape}
& B^{\rm loc.}_{\Phip} (  k_1 , k_2 , k_3 ) = 2 \Delta_\Phi^2 \left( \frac{1}{k_1^3 k_2^3} +\frac{1}{k_1^3 k_3^3}+\frac{1}{k_2^3 k_3^3}   \right) \ ,  \\ 
& B^{\rm equil.}_{\Phip} ( k_1 , k_2 , k_3) = 6 \Delta_\Phi^2  \left(- \frac{1}{k_1^3 k_2^3}  -  \frac{1}{k_1^3 k_3^3}  -  \frac{1}{k_2^3 k_3^3}  - \frac{2}{k_1^2 k_2^2 k_3^2} + \left(  \frac{1}{k_1 k_2^2 k_3^3}  + 5 \text{ perms.} \right)  \right)  \ ,  \\
& B^{\rm orth.}_{\Phip}  ( k_1 , k_2 , k_3 ) = 6 \Delta_\Phi^2 \left(  (1 + p) \frac{\Delta ( k_1 , k_2 , k_3 ) }{k_1^3 k_2^3  k_3^3}   - p \frac{\Gamma (k_1 , k_2 , k_3 )^3 }{k_1^4 k_2^4 k_3^4}  \right) \ ,
\end{split}
\end{align}
with $p = 8.52$, 
\be
\Delta ( k_1  , k_2 , k_3 ) = (-k_1 + k_2 + k_3) (k_1 - k_2 + k_3) ( k_1 + k_2 - k_3)   \ , 
\ee
and
\be
\Gamma ( k_1 , k_2 , k_3 ) = \frac{2}{3} (k_1 k_2 + k_2 k_3 + k_3 k_1 ) - \frac{1}{3} (k_1^2 + k_2^2  + k_3^2 ) \ ,
\ee
and where the primordial power spectrum is given by
\be\label{eq:primordial_power}
P_{\Phi_{\rm p}} ( k ) = \frac{\Delta_\Phi }{k^3} \left(  \frac{k}{k_0} \right)^{n_s-1}  \ ,
\ee
$n_s$ is the scalar tilt, and $k_0$ is the pivot scale. 

For example, in the context of the Effective Field Theory of Inflation~\cite{Cheung:2007st}, the signal produced  by the two leading interactions for the Goldstone boson of time translations, $\pi$, that parametrizes the inflaton fluctuations, $\dot\pi^3$ and $\dot\pi(\partial_i\pi)^2$, is well described by a linear combination of $\fnl^{\rm equil.}$ and $\fnl^{\rm orth.}$~\cite{Senatore:2009gt}. Many multifield inflationary  models are well described by a combination that also includes $\fnl^{\rm loc.}$~\cite{Bernardeau:2002jy,Lyth:2002my,Zaldarriaga:2003my,Babich:2004gb,Senatore:2010wk}. 
Additionally, quasi-single field inflationary models~\cite{Chen:2009zp,Sefusatti:2012ye}, where particles with mass of order of the inflationary Hubble rate are present, produce signals that in some kinematical limits are not well described by a combination of the three shapes above, but nevertheless have most of the signal to noise well reproduced by such a combination of shapes of three-point functions. In this paper, we develop the methods to analyze the three shapes above on the BOSS DR12 galaxy sample {based on the power spectrum and bispectrum in the EFTofLSS at one loop}. Our results are summarized in Fig.~\ref{fig:fnl_summary}.

The paper is organized as follows.  {We present in~\secref{sec:theory} the predictions of the power spectrum and bispectrum at one loop in the presence of non-Gaussianities. 
A rather new aspect of galaxy-clustering analysis from the EFTofLSS is highlighted in~\secref{ppsec} based on considerations of internal theoretical consistencies. 
We then present our main results and conclusions in~\secref{resutltssec}. 
Many details concerning the analysis are relegated to appendices. 
A discussion on non-Gaussian bias parameters and their impact on the constraints of the local shape is given in~\appref{ngbiasapp}. 
Technical aspects of our inference setup are provided in~\appref{sec:likelihood}. 
Finally in~\appref{sec:plots}, we provide additional triangle plots of various analyses performed in this work.}

\paragraph{Data sets:} We analyze the power spectrum and the bispectrum of the SDSS-III BOSS DR12 galaxy sample~\cite{BOSS:2016wmc}. 
The power spectrum, window functions, and bispectrum, are measured from BOSS catalogs DR12 combined CMASS-LOWZ~\cite{Reid:2015gra}.\footnote{publicly available at \href{https://data.sdss.org/sas/dr12/boss/lss/}{https://data.sdss.org/sas/dr12/boss/lss/}}
The covariances are estimated from the 2048 patchy mocks~\cite{Kitaura:2015uqa}. 
To test our analysis pipeline, we analyze the mean over the 84 Nseries `cut-sky' mocks, which are full $N$-body simulations populated with a Halo Occupation Distribution model and selection function similar to the one of BOSS~\cite{BOSS:2016wmc}.\footnote{Catalog available at \href{https://www.ub.edu/bispectrum/page11.html}{https://www.ub.edu/bispectrum/page11.html}}
For the PS, we fit the monopole and quadrupole, up to $\kmax = 0.23 \hinvMpc$, as determined in~\cite{DAmico:2019fhj,Colas:2019ret,DAmico:2020kxu,Zhang:2021yna}. 
We jointly fit the BOSS DR12 bispectrum monopole, up to $\kmax = 0.23 \hinvMpc$, as we explain in the following. 
More details on our measurements can be found in App.~\ref{sec:likelihood}.

\paragraph{Public codes:}   The predictions for the full shape of the galaxy power spectrum in the EFTofLSS are obtained using \code{PyBird}: Python code for Biased tracers in Redshift space~\cite{DAmico:2020kxu}.\footnote{\href{https://github.com/pierrexyz/pybird}{https://github.com/pierrexyz/pybird}}
The linear power spectra were computed with the \code{CLASS} Boltzmann code v2.7~\cite{Blas_2011}.\footnote{ \href{http://class-code.net}{http://class-code.net}}
The posteriors were sampled using the \code{MontePython} v3.3 cosmological parameter inference code~\cite{Brinckmann:2018cvx, Audren:2012wb}.\footnote{ \href{https://github.com/brinckmann/montepython\_public}{https://github.com/brinckmann/montepython\_public}}
The plots have been obtained using the \code{GetDist} package~\cite{Lewis:2019xzd}. 
The power spectrum multipoles and the bispectrum monopole are measured using \code{Rustico}~\cite{Gil-Marin:2015sqa}.\footnote{\href{https://github.com/hectorgil/Rustico}{https://github.com/hectorgil/Rustico} }  The PS window functions are measured as described in~\cite{Beutler:2018vpe} using \code{fkpwin}~\cite{Simon:2022adh}\footnote{\href{https://github.com/pierrexyz/fkpwin}{https://github.com/pierrexyz/fkpwin}} based on \code{nbodykit}~\cite{Hand:2017pqn}.\footnote{\href{https://github.com/bccp/nbodykit}{https://github.com/bccp/nbodykit}}

%
%
%
\section{EFTofLSS with non-Gaussianity\label{sec:theory}}

\subsection{Biased tracers in redshift space} \label{btrssapp}
We first focus on biased tracers with Gaussian initial conditions, and introduce non-Gaussianities in the next subsection.    The transformation of the halo overdensity $\delta_h$ (in position space) to redshift space is given by 
\begin{align}
\begin{split} \label{rsssbias}
\delta_{r,h} ( \xvec ) &  = \delta_h ( \xvec ) - \frac{\hat{z}^i \hat{z}^j}{aH } \partial_i \left( ( 1 + \delta_h ) v^j \right) + \frac{\hat{z}^i \hat{z}^j \hat{z}^k \hat{z}^l}{2 (a H)^2  } \partial_i \partial_j ( ( 1 + \delta_h ) v^k  v^l ) \\
& - \frac{\prod_{a=1}^6 \hat{z}^{i_a} }{3! (a H)^3 } \partial_{i_1} \partial_{i_2} \partial_{i_3} (( 1 + \delta_h ) v^{i_4} v^{i_5} v^{i_6} ) + \frac{\prod_{a=1}^8 \hat{z}^{i_a} }{4! (a H)^4} \partial_{i_1} \partial_{i_2} \partial_{i_3}\partial_{i_4}  (  v^{i_5 }  v^{i_6} v^{i_7} v^{i_8} )  + \dots \ .
\end{split}
\end{align}
Next, we perturbatively expand the halo overdensity in redshift space 
\be
\delta_{r,h} ( \kvec ; \hat z ) = \sum_n  \delta^{(n)}_{r,h} ( \kvec ; \hat z) \ ,
\ee
where $\hat z$ is the line-of-sight direction.  For our analysis, we consider the one-loop power spectrum and the {one-loop} bispectrum, which means that we need the halo overdensity in redshift space up to {fourth} order.
The solutions (ignoring EFT counterterms for now, we will return to them later) can be written as\footnote{We have introduced the following notation
\be
\int_{\kvec_1 , \dots , \kvec_n} \equiv \int \frac{d^3 k_1 }{(2 \pi)^3} \cdots \frac{d^3 k_n}{(2 \pi)^3 } \ ,  \quad \int_{\kvec_1 , \dots , \kvec_n}^{\kvec} \equiv \int_{\kvec_1 , \dots , \kvec_n} ( 2 \pi)^3 \delta_D ( \kvec - \sum_{i = 1}^n \kvec_i )  \ .
\ee
{Additionally, $\delta^{(1)}$ and $P_{11}$ are the dark-matter linear overdensity field and power spectrum, respectively.}}
\begin{align}
\begin{split} \label{halorsskernels}
\delta_{r,h}^{(1)} ( \kvec ; \hat z ) & = K_1^{r,h} ( \kvec ; \hat z ) \delta^{(1)} ( \kvec ) \ ,  \\
\delta_{r,h}^{(n)} ( \kvec  ; \hat z ) & =  \int_{\kvec_1 , \dots , \kvec_n}^{\kvec } K_n^{r,h} ( \kvec_1 , \dots , \kvec_n  ; \hat z ) \delta^{(1)} ( \kvec_1 ) \cdots \delta^{(1)}(\kvec_n) \ , 
\end{split}
\end{align}
for $n \geq 2$, which defines the $n$-th order symmetric halo kernels $K_{n}^{r,h}$.  The kernels $K_{1,\dots,4}^{r,h}$  
depend on the bias parameters $\{ b_i \}$ for $i = 1, \dots, 15$ in the following way
\begin{align}
\begin{split}
& K_{1}^{r,h} [ b_1 ] \ , \quad  K_{2}^{r,h} [ b_1 , b_2 , b_5 ] \ , \quad  K_{3}^{r,h} [ b_1 , b_2 , b_3, b_5 , b_6, b_8, b_{10}  ]  \andd  K_4^{r,h} [ b_1, \dots , b_{15}]  \  . 
\end{split}
\end{align}
For example,
\be
 K_1^{r,h} ( \kvec ; \hat z )  =  b_1 + f (\hat k \cdot \hat z )^2    \ , 
\ee
is the Kaiser result, and explicit expressions for $K_2^{r,h} $ and $K_3^{r,h} $ can be found in \cite{Perko:2016puo}, while the expression for $K_4^{r,h}$ is available in \cite{DAmico:2022ukl} and is a conceptually straightforward extension to next order of the EFT bias expansion.   {For reference, the second-order kernel, not in redshift space (the transformation to which can be read directly from \eqn{rsssbias}), is given by}
\be
K_2^h ( \kvec_1 , \kvec_2 ) = b_1 \frac{\kvec_1 \cdot \kvec_2}{2} \left(\frac{1}{k_1^2 } + \frac{1}{k_2^2}  \right)  + b_2 \left( \frac{5}{7} + \frac{2 (\kvec_1 \cdot \kvec_2)^2}{ 7 k_1^2 k_2^2 } \right)   + b_5 \ . 
\ee
{All explicit expressions for the perturbative kernels in redshift space up to $n = 4$ can be found in the ancillary Mathematica file associated to \cite{DAmico:2022ukl}, which also uses the same notation as this paper.  }

The observables that concern us here are the one-loop power spectrum and the one-loop bispectrum.  The one-loop power spectrum is $P_{11}^{r,h}+P_{1\text{-loop}}^{r,h} $, with $P_{11}^{r,h}$ the tree-level power spectrum:
\be 
P_{11}^{r,h} ( k , \hat k \cdot \hat z) = (b_1 + f (\hat k \cdot \hat z )^2  )^2  P_{11} ( k )  \ , 
\ee
and $P_{1\text{-loop}}^{r,h} \equiv P_{22}^{r,h} + P_{13}^{r,h}$, the one-loop contribution, with 
\begin{align}
\begin{split} \label{loopexpressionsrssbias}
&P_{22}^{r,h} ( k, \hat k \cdot \hat z  )  = 2 \int_{\qvec} K_2^{r,h} ( \qvec , \kvec - \qvec ; \hat z )^2 P_{11} ( q ) P_{11}( | \kvec - \qvec|) \ , \\
& P_{13}^{r,h}  (k, \hat k \cdot \hat z)  = 6 P_{11} ( k )K_1^{r,h} ( \kvec ; \hat z )  \int_{\qvec} K_3^{r,h} ( \qvec , - \qvec , \kvec ;  \hat z ) P_{11}(q) \ .
\end{split}
\end{align}
The {one-loop} bispectrum is given by $B_{211}^{r,h}+B_{1\text{-loop}}^{r,h}$, with the tree-level bispectrum:
\be
B_{211}^{r,h} = 2  K_1^{r,h} ( \kvec_1 ; \hat z )  K_1^{r,h} ( \kvec_2  ; \hat z )  K_2^{r,h} ( -\kvec_1, -\kvec_2 ;  \hat z ) P_{11} ( k_1 ) P_{11} ( k_2 ) + \text{ 2 perms.}  \ ,
\ee
(we have dropped the argument $( k_1 , k_2 , k_3 , \hat k_1 \cdot \hat z , \hat k_2 \cdot \hat z ) $ on the bispectrum to remove clutter), and the one-loop contribution:
\be
B_{1\text{-loop}}^{r,h}= B^{r,h}_{222} + B_{321}^{r,h,(I)} + B_{321}^{r,h,(II)}  + B_{411}^{r,h} \ , 
\ee
with
\begin{align}
 \label{bispexpressionsrssbias}
& B^{r,h}_{222}  = 8 \int_{\qvec} P_{11}(q) P_{11}(| \kvec_2 - \qvec| ) P_{11} ( |\kvec_1 + \qvec|) \nonumber \\
& \hspace{1.5in} \times K_2^{r,h} ( - \qvec , \kvec_1 + \qvec ; \hat z) K^{r,h}_2 ( \kvec_1 + \qvec , \kvec_2 - \qvec ; \hat z ) K^{r,h}_2 ( \kvec_2 - \qvec , \qvec ;  \hat z) \ , \nonumber   \\
& B_{321}^{r,h,(I)}  = 6 P_{11}(k_1) K_1^{r,h} ( \kvec_1 ;  \hat z)  \int_{\qvec} P_{11}(q) P_{11}(| \kvec_2 - \qvec|)      \\
& \hspace{1.5in} \times K_3^{r,h} ( - \qvec , - \kvec_2 + \qvec , - \kvec_1 ; \hat z ) K^{r,h}_2 ( \qvec , \kvec_2 - \qvec  ;  \hat z )  + \text{ 5 perms.}\ ,   \nonumber   \\
& B_{321}^{r,h,(II)}  = 6 P_{11}(k_1 ) P_{11}(k_2) K_1^{r,h} ( \kvec_1  ; \hat z) K_2^{r,h} ( \kvec_1 , \kvec_2 ;  \hat z ) \int_{\qvec} P_{11}(q) K_3^{r,h} ( \kvec_2 , \qvec , - \qvec ; \hat z ) + \text{ 5 perms.} \ ,  \nonumber \\ 
& B_{411}^{r,h}  = 12 P_{11} ( k_1 ) P_{11}(k_2) K_1^{r,h} ( \kvec_1 ; \hat z) K_1^{r,h} ( \kvec_2 ; \hat z)  \int_{\qvec} P_{11}(q) K_4^{r,h} ( \qvec , - \qvec , - \kvec_1 , - \kvec_2 ; \hat z ) + \text{ 2 perms.}  \ . \nonumber
\end{align}

Next we turn to the counterterms that renormalize the one-loop power spectrum.  Since we are working directly with biased tracers, we can introduce the counterterms directly into \eqn{rsssbias}, as in {\cite{Perko:2016puo}} for example.  The counterterms come from two sources.  The first are response terms, which give a contribution:
\be
P_{13}^{r,h,ct} ( k , \hat k \cdot \hat z ) = 2 K_1^{r,h}( \kvec ; \hat z ) K_1^{r,h,ct} ( - \kvec ; \hat z ) P_{11} ( k ) \  ,
\ee
where
\be
K_1^{r,h,ct} ( \kvec ; \hat z)  = \frac{k^2}{k_{\rm R}^2} \left(  - \frac{k_{\rm R}^2}{\km^2} c_{1}^h +  f (\hat k \cdot \hat z)^2 c_{1}^\pi - \half f^2 (\hat k \cdot \hat z)^4 c_{1}^{\pi v} - \half f^2 (\hat k \cdot \hat z)^2 c_{3}^{\pi v} \right) \ ,
\ee
{and $k_{\rm M}$ is the non-linear scale associated with dark-matter clustering and halo formation, and $k_{\rm R}$ is the non-linear scale associated with redshift space.}
We also have the stochastic terms, which give a contribution:
\be
P_{22}^{r,h,\epsilon} ( k , \hat k \cdot \hat z) = \frac{1}{\bar n} \left( c_1^{\rm St}  +c^{\rm St}_2  \frac{k^2}{k_{\rm M}^2}  + c^{\rm St}_3 \frac{k^2}{k_{\rm M}^2} f ( \hat k \cdot \hat z)^2    \right)  \ . 
\ee
The one-loop bispectrum is renormalized by response terms in 
\be
B_{411}^{r,h,ct}  = 2 P_{11} ( k_1 ) P_{11} ( k_2 )K_1^{r,h}(\kvec_1 ; \hat z ) K_1^{r,h} (\kvec_2 ; \hat z )   K_{2}^{r,h,ct} ( - \kvec_1 , - \kvec_2 ; \hat z  )  + \text{ 2 perms.} \ ,
\ee
where $K_{2}^{r,h,ct}$ is given in \cite{DAmico:2022ukl}, and
\be
B_{321}^{r,h,(II),ct}  = 2 P_{11} ( k_1 ) P_{11} ( k_2 ) K_{1}^{r,h,ct} ( \kvec_1 ; \hat z ) K_1^{r,h} ( \kvec_2 ; \hat z ) K^{r,h}_2 ( - \kvec_1 , - \kvec_2 ; \hat z )  + \text{ 5 perms.} \ .
\ee
In a similar way, the stochastic contributions $B_{321}^{r,h,(I),\epsilon} $ and $B_{222}^{r,h,\epsilon}$ are given in \cite{DAmico:2022ukl} and can be found by a straightforward application of EFTofLSS principles.  {Numerically, in our analysis,} we use {$\bar n = 4 \cdot 10^{-4} \, (\Mpcinvh)^{3}$}, $k_{\rm M} = 0.7 \, h/\text{Mpc}$, and {$k_{\rm R }  = 0.25 \, h/\text{Mpc}$, and we have $f(z = 0.57) =0.78 $ for our cosmology}.

{In summary, the final expressions that we use for the total one-loop power spectrum $P^{r,h}|_1$ and total one-loop bispectrum $B^{r,h} |_1$ without non-Gaussianities are}
\begin{align}
\begin{split}
P^{r,h} |_1 & = P_{11}^{r,h} + P_{1\text{-loop}}^{r,h} + P_{13}^{r,h,ct} + P_{22}^{r,h,\epsilon}  \\
B^{r,h} |_1& = B^{r,h}_{211} + B^{r,h}_{1\text{-loop}} + B_{411}^{r,h,ct} + B_{321}^{r,h,(II),ct} + B_{321}^{r,h,(I),\epsilon} + B_{222}^{r,h,\epsilon} \ . 
\end{split}
\end{align}

{In redshift space, we analyze the power-spectrum monopole and quadrupole, and the bispectrum monopole.  The power-spectrum multipoles are given by 
\be
P_\ell^{r,h} ( k ) = \frac{2 \ell + 1}{2} \int_{-1}^{1} d \mu \, \mathcal{P}_\ell ( \mu ) P^{r,h} ( k , \mu) \ , 
\ee
where $\mathcal{P}_\ell$ are the Legendre polynomials,  and the bispectrum monopole is given simply by the average over the redshift space angles \cite{Scoccimarro:1999ed, Scoccimarro:2015bla, Gil-Marin:2016wya}\footnote{We have corrected a factor of $1/(4 \pi)$ in Eq. (14) of \cite{Gil-Marin:2016wya}.}
\be
B_0^{r,h} ( k_1 , k_2 , k_3 ) = \frac{1}{4 \pi } \int_{-1}^1 d \mu_1 \int_{0}^{2 \pi} d \phi  \, B^{r,h} ( k_1 , k_2 , k_3, \mu_1 , \mu_2 )  \ ,
\ee 
where $\mu_2 \equiv \mu_1 \hat k_1 \cdot \hat k_2 - \sqrt{1 - \mu_1^2} \sqrt{1- (\hat k_1 \cdot \hat k_2)^2} \cos \phi $. }


%
\subsection{Non-Gaussianity}
We now pass to introduce the modifications needed if the primordial fluctuations are non-Gaussian. We describe non-Gaussianity in the primordial potential, $\Phi_{\rm p}^{\rm NG}$, in terms of a Gaussian auxiliary potential $\Phi_{\rm p}^{\rm G}$ as:
\be \label{phing}
\Phi^{\rm NG}_{\rm p} ( \kvec ) = \Phip^{\rm G} ( \kvec ) + f_{\rm NL} \int_{\kvec_1 , \kvec_2}^{\kvec} W( \kvec_1 , \kvec_2 ) \Phip^{\rm G} ( \kvec_1 ) \Phip^{\rm G} ( \kvec_2 )  \ ,
\ee
which gives to leading order in $\fnl$ the primordial bispectrum:
\be
B_{\Phip} ( k_1 , k_2 , k_3 ) = 2 \fnl \left( W( \kvec_1 , \kvec_2) P_{\Phip} ( k_1 ) P_{\Phip} ( k_2) + 2\text{ perms.} \right)   \ ,
\ee
where $P_{\Phip} $ is the primordial power spectrum \eqn{eq:primordial_power}.
A choice of $W$ that gives the desired bispectrum at leading order in $\fnl$ is 
\be
W(\kvec_1 , \kvec_2 ) = \frac{1}{6}\frac{B_{\Phip} ( k_1 , k_2 , | \kvec_1 + \kvec_2| )  }{P_{\Phip}(k_1) P_{\Phip} ( k_2 ) }  \ . 
\ee
Then, using the transfer function $T_\alpha$ defined by
\be
\delta^{(1)} ( \kvec , a ) = T_\alpha ( k , a )  \Phi_{\rm p} ( \kvec )   \ , 
\ee
we can convert  \eqn{phing} to 
\be \label{deltang}
\delta^{(1)}_{\rm NG} ( \kvec,a ) = \delta^{(1)} ( \kvec ,a  ) + \fnl \int_{\kvec_1 , \kvec_2}^{\kvec} W( \kvec_1 , \kvec_2 ) \frac{T_\alpha ( | \kvec_1 + \kvec_2| ,a)}{T_\alpha( k_1, a) T_\alpha ( k_2 , a) } \delta^{(1)} ( \kvec_1 , a ) \delta^{(1)} ( \kvec_2 , a )  \ ,
\ee
where $\delta^{(1)}$ is a Gaussian field.  

There are two effects from non-Gaussianity on the perturbative expansion.  The first comes from replacing all of the $\delta^{(1)}$ initial conditions in \eqn{halorsskernels} with the non-Gaussian initial conditions \eqn{deltang}.  At the level to which we work, this simply produces a shift in the second-order kernel
\be \label{k2kerexp}
\delta K^{h,r}_2 ( \kvec_1 , \kvec_2 ;  \hat z ) = \fnl K^{h,r}_1 ( \kvec_1 + \kvec_2 ; \hat z)  \frac{T_\alpha ( | \kvec_1 + \kvec_2 | , a )}{T_\alpha ( k_1 , a ) T_\alpha ( k_2 , a ) } W( \kvec_1 , \kvec_2 )  \ ,
\ee
which therefore, at lowest order in $\fnl$ and $k/\knl$ at which we work, modifies the bispectrum.  

The second effect of $\fnl$ comes from allowing new terms (which appear non-local because of initial long-range correlations {\cite{Dalal:2007cu,Slosar:2008hx, Verde:2009hy, Schmidt:2010gw, Angulo:2015eqa,Assassi:2015jqa,  Lewandowski:2015ziq}}) in the bias expansion.  In order to define the new counterterms, and because of the smoothing procedure used to define the EFT, we look at the squeezed limit of the last term in \eqn{deltang}.  This defines a new long-wavelength field 
\be
\tilde \phi ( \kvec_L , a ) \equiv \frac{W_{\rm SL} ( \knl , \kvec_L) }{T_\alpha ( k_L , a )} \delta^{(1)} ( \kvec_L , a ) \ ,
\ee
which can now be used to build counterterms.  The function $W_{\rm SL}$ is defined by the squeezed limit of $W( \kvec_S , \kvec_L)$ for $k_L \ll k_S$, with an angle average used for $W_{\rm SL}^{\rm equil.}$.(\footnote{We neglect effects due to the angle $\hat k_S\cdot \hat k_L$ \cite{Assassi:2015jqa, Lewandowski:2015ziq}. })  Explicitly, we have
\be
W_{\rm SL} ( \knl , \kvec_L ) = w_\beta \left( \frac{k_L}{\knl} \right)^\beta  \ , 
\ee
with 
\be
(\beta , w_\beta )_{\rm loc} = (0, 2 / 3) \ , \quad (\beta , w_\beta )_{\rm equil.} = (2, 4 / 3) \andd  (\beta , w_\beta )_{\rm orth.} = (2, 4(9-7p)/27) \ .
\ee
{where $p = 8.52$}. Thus, the new terms that we can write in the renormalized halo density are
\begin{align}
&[\delta_{r,h}  ( \kvec ; \hat z )  ]^{\fnl}   =  b^{\fnl}_1 \fnl \tilde \phi  ( \kvec , a_{\rm in})  \\
&\hspace{1in}  +  \fnl  \int_{\kvec_1 , \kvec_2}^{\kvec}  \Bigg(   b_1^{\fnl} \left(    \frac{\kvec_1 \cdot \kvec_2}{k_2^2}  + f \hat z \cdot (\kvec_1 + \kvec_2) \frac{\hat z \cdot \kvec_2}{k_2^2}  \right)   +  b_2^{\fnl}    \Bigg)   \tilde \phi ( \kvec_1 , a_{\rm in} ) \delta^{(1)} ( \kvec_2 )  \nonumber \ .
\end{align}
where the first term in the second line is the flow term associated to the operator in $b^{\fnl}_1$ \cite{Senatore:2014eva} and $f$ is the linear growth rate.  We give details on our specific choices for the expressions for the non-Gaussian biases $b_{1}^{\fnl}$ and $b_2^{\fnl}$ in \appref{ngbiasapp}.

%
%
%
\section{Perturbativity prior}  \label{ppsec}

{While we relegate to App.~\ref{sec:likelihood} the details on our inference setup (data, likelihood, standard prior, and posterior sampling) and on additional modeling for the BOSS data, we now introduce a rather new interesting aspect of our analysis, which is a theoretically justified restriction on the ranges of the EFT parameters.}  Since we work in perturbation theory, we rely on the fact that higher loop contributions that we do not include are sufficiently smaller than the error bars of the data and of the overall size that is theoretically expected. Thus, in the Markov chain Monte Carlo (MCMC) analysis, we want to penalize values of EFT parameters that lead to large loop contributions, since these correspond to configurations that are not perturbative or physical.  {Therefore, we further impose a prior on the size of the loop using considerations from the perturbative nature of our predictions, following \cite{Braganca:2023pcp}.  Because it will capture most of the effect that we seek, we work in real space, i.e. with $f=0$, and we use the superscript $``h"$ to represent quantities in \secref{sec:theory} with $f=0$.  We stop our analysis at a $\kmax$ such that the signal-to-noise of the neglected two-loop terms (which are the lowest order that we neglect) is a fraction of the total signal-to-noise (estimated as $N_{\rm bins}^P$ and $N_{\rm bins}^B$ for the power spectrum and bispectrum respectively), 
\be \label{p2loopsum}
\sum_{i \in \text{bins}_P}^{\kmax}  \left( \frac{P_{2\text{-loop}}^h ( k^i ) }{\sigma_P^{\rm data} ( i )}  \right)^2 \approx X_{P}^2 N_{\rm bins}^P \andd
\sum_{i \in \text{bins}_B}^{\kmax}  \left( \frac{B_{2\text{-loop}}^h ( k^i_1, k_2^i , k_3^i ) }{\sigma_B^{\rm data} ( i )}  \right)^2 \approx X_{B}^2 N_{\rm bins}^B \ , 
\ee
where $k^i_a$ is the value of $k_a$ in bin $i$, $\sigma^{\rm data}_{P,B} (i)$ are the error bars on the data (we assume a diagonal covariance for our estimates here), $N_{\rm bins}^P$ is the number of power spectrum bins, $N_{\rm bins}^B$ is the number of bispectrum bins, the sums are over all bins with a maximum wavenumber of $\kmax$, and $X_P$ and $X_B$ are the tolerances for the size of neglected contributions in each bin for the power spectrum and bispectrum, respectively, in units of the error bars of the bin.  {Given that we tolerate a theoretical error of $\sigma^{\rm data}/3$, we expect $X_P $ and $X_B$ to be around 1/3, up to an order one number. In practice, in order to obtain unbiased results,} we use $X_P = 1/3$ and $X_B = 1/2$ in our analysis.  Good proxies for the sizes of the neglected terms are
\begin{equation}\label{eq:twoloop}
|P^h_{\text{2-loop}}| \sim ( P^h_{\text{1-loop}})^2 / P^h_{11} \andd  |B^h_{\text{2-loop}} | \sim | B^h_{\text{1-loop}} \, P^h_{\text{1-loop}} |  / P^h_{11} \ . 
\end{equation}

Next, the expected scalings of the one-loop power spectrum and counterterm are well approximated by (see \cite{Braganca:2023pcp})
\begin{align}
S^P_{\text{1-loop}}(k) &\sim b_1^2 P_{11}(k) \left(\frac{k}{\knl}\right)^{3+n(k)}  \ , \\
S^P_{\rm ct}(k) & \sim 2 b_1 P_{11}(k) \left(\frac{k}{\knl}\right)^{2} \ ,
\end{align}
where $P_{11}$ is the linear matter power spectrum, and $n(k) \equiv d\log P_{11}/ d\log k$.  
Similarly, for the one-loop bispectrum, the expected scalings are
\begin{align}
S^B_{\text{1-loop}}(k_1, k_2, k_3) & \sim B^h_{211}(k_1,k_2,k_3) \sum_{i=1}^3 \left(\frac{k_i}{\knl}\right)^{3+n(k_i)} \ , \\
S^B_{\rm ct}(k_1, k_2, k_3) & \sim 2 b_1^2 \left( P_{11}(k_1) P_{11}(k_2) \left(\frac{k_3}{\knl}\right)^{2} + \text{2 cyc.} \right) \ .
\end{align}
By defining $S^P \propto \max(S^P_{\text{1-loop}}, S^P_{\rm ct})$ and $S^B \propto \max (|S^B_{\text{1-loop}}|, S^B_{\rm ct})$ normalized so that $S^P(k_{\rm max}) = 1$ and $ S^B(k_{\rm max}) = 1$, the expected sizes for the one-loop contributions in our predictions are, for all wavenumbers,
\begin{equation} \label{oneloopscaling}
\sigma^{\rm P.P.}_P (k) \sim S^P(k) P_{\text{1-loop}}^{k_{\rm max}} \ , \qquad \sigma^{\rm P.P.}_B (k_1, k_2, k_3) \sim S^B(k_1, k_2, k_3) B_{\text{1-loop}}^{k_{\rm max}}  \ .
\end{equation}

Now, we can estimate the values of $P_{\text{1-loop}}^{k_{\rm max}} $ and $B_{\text{1-loop}}^{k_{\rm max}} $ where we stop our analysis by plugging in the two-loop estimates \eqn{eq:twoloop} and the one-loop scaling relations \eqn{oneloopscaling} into \eqn{p2loopsum}.  We find
\be
P_{1\text{-loop}}^{\kmax} = X_P^{1/2} (N_{\rm bins}^P)^{1/4} \left[ \sum_{i \in \text{bins}_P}  \frac{S^P(k^i)^4}{P_{11}^h ( k^i )^2 \sigma_P^{\rm data} ( i )^2 }  \right]^{-1/4} \ , 
\ee 
and
\begin{align}
B_{\text{1-loop}}^{k_{\rm max}}   = X_B (N_{\rm bins}^B)^{1/2} \left( \sum_{i \in \text{bins}_B}  \left( \frac{S^B(k_1^i , k_2^i , k_3^i)}{\sigma_B^{\rm data} ( i ) } \frac{1}{3} \sum_{a=1}^3 \frac{ S^P ( k_a^i) P_{1\text{-loop}}^{\kmax}   }{P_{11}^h ( k_a^i ) } \right)^2  \right)^{-1/2}   \  . 
\end{align}

{Now, by choosing to penalize the likelihood when the size of the one-loop contribution exceeds its expectation in perturbation theory, potential unphysical predictions will not contribute to the posterior. 
We thus add the following `perturbativity' prior to $-2 \log \mathcal{L}$: 
\begin{equation}\label{eq:pertP}
\mathcal{P}_P = \frac{1}{2 N^P_{\rm bins}} \sum_{i \in \text{bins}_P} \left( \frac{P^h_{\text{1-loop}}(k_i)}{\sigma^{\rm P.P.}_P (k_i) }\right)^2 \ ,
\end{equation}
when analyzing the power spectrum at one-loop, and
\begin{equation}\label{eq:pertB}
\mathcal{P}_B = \frac{1}{2 N^B_{\rm bins}} \sum_{i \in \text{bins}_B} \left( \frac{B^h_{\text{1-loop}}(k_1^i, k_2^i, k_3^i)}{\sigma^{\rm P.P.}_B (k_1^i, k_2^i, k_3^i)} \right)^2 \ ,
\end{equation}
when analyzing the bispectrum at one-loop. This ensures that a global departure of the size of the one-loop with respect to its expectation would penalize $-2 \log \mathcal{L}$ by $\mathcal{O}(1)$, independently of the number of bins. }

%
%
%

\section{Results and conclusions} \label{resutltssec}

\paragraph{Main results:} We perform the analysis of the power spectrum monopole and quadrupole as well as of the bispectrum monopole of the BOSS DR12 galaxy-clustering data, by fixing $\Lambda$CDM parameters to Planck preferred values~\cite{Planck:2018vyg}, and by scanning over $\fnl$. We use the EFTofLSS prediction at one-loop order {\it both} for the power spectrum {\it and} for the bispectrum. This allows us to reach relatively high wavenumber in both statistics, {$\kmax=0.23 \hinvMpc$} {for the power spectrum and the bispectrum}, and therefore to extract much information from the data, {as we will describe below}. The development of a pipeline that allows us to analyze the one-loop bispectrum predicted by the EFTofLSS  has required much theoretical work, and the explanation  of such techniques are presented in~\cite{DAmico:2022osl,DAmico:2022ukl,Anastasiou:2022udy}. Here, instead, we will just give the essential details by focussing on the application of such techniques to the analysis of the BOSS data for primordial non-Gaussianities.

We first calibrate the model against the set of N-series simulations.  These are simulations run with Gaussian initial conditions, such that the 84 N-series boxes represent a volume of approximately $\sim 50$ times the total BOSS volume.  In order to estimate our theoretical error, we run our chains on data which is the average over all of the 84 N-series boxes, {using the covariance corresponding to the total volume}.  
We find
\bea \label{simsresultseq}
&& 98  <\fnl^{\rm equil., sims}<  226 \, , \quad {\rm or} \quad\fnl^{\rm equil., sims}=162  \pm 64\ ,  \quad {\rm at} \  68\%\  {\rm CL} \ ,\\  \label{simsresultsor}
&& - 52 <\fnl^{\rm orth., sims}< -12 \,,\quad {\rm or} \quad\fnl^{\rm orth., sims}=  - 32  \pm 20 \ ,\quad {\rm at} \  68\%\  {\rm CL} \ , \\ \label{simsresultslo}
&& -1 <\fnl^{\rm loc., sims}< 7 \,,\quad {\rm or} \quad\fnl^{\rm loc., sims}= 3 \pm 4\ , \quad {\rm at} \  68\%\  {\rm CL}\ .
\eea
Using the errors found above, and the errors that we will find on the BOSS data in Eqs.~(\ref{fnleqconst})~-~(\ref{fnllocconst}), we can estimate the theoretical systematic error in our analysis.  
First, let us in general write $f_{\rm NL}^{s{\rm , sims}} = \mu_{s}^{\rm sims} \pm \sigma_{s}^{\rm sims}$, and $f_{\rm NL}^{s{\rm , BOSS}} = \mu_{s}^{\rm BOSS} \pm \sigma_{s}^{\rm BOSS}$, where $\mu$ is the mean of the measurement, $\sigma$ is the $1\sigma$ error, and $s$ stands for the shape $ s \in \{\text{equil.}, \text{orth.} , \text{loc.} \}$.   {Next, we declare that the minimum theoretical error that we can detect is given by $\sigma_s^{\rm sims}$, the $68\%$-confidence intervals measured on the simulations. }  Then, it is meaningful to estimate the theoretical error of the EFTofLSS prediction, $\sigma^{\rm th. sys.}_{s}$, as  the distance of the mean of the distribution to the zero value of $\fnl$ minus $\sigma_s^{\rm sims}$, or $\sigma_s^{\rm sims}$ if this number is negative.  In equations, this is
\be
\sigma^{\rm th. sys.}_{s} \lesssim   | \mu_{s}^{\rm sims} | - \sigma_s^{\rm sims}\ ,
 \ee
or $\sigma^{\rm th. sys.}_{s} \lesssim \sigma_s^{\rm sims}$ if this number is negative.  In units of the standard deviation that we will find in the data, we therefore can estimate  {${\sigma^{\rm th. sys.}_{\rm equil.}}\lesssim 0.34 \sigma^{\rm BOSS}_{\rm equil.}$}, ${\sigma^{\rm th. sys.}_{\rm orth.}} {\lesssim} 0.16 \sigma^{\rm BOSS}_{\rm orth.} $, and ${\sigma^{\rm th. sys}_{\rm loc.}}{\lesssim} {0.12} \sigma^{\rm BOSS}_{\rm loc.}$.   This allows us to conclude that the {errors associated to the modeling and the analysis methods are safely negligible.  Contour plots associated to these analyses are presented in~App.~\ref{sec:plots}.

Having found satisfactory results against simulations, we move to analyze the BOSS data. We find
\bea \label{fnleqconst}
 && {- 85  <\fnl^{\rm equil.,\, BOSS}<  499 \,, \quad {\rm or} \quad\fnl^{\rm equil.,\, BOSS}= 207  \pm 292\ ,   \quad {\rm at} \  68\%\  {\rm CL} }  \ ,\\
&& -141  <\fnl^{\rm orth.,\, BOSS}< 5 \,, \quad {\rm or} \quad\fnl^{\rm orth.,\, BOSS}= -68  \pm 73 \ ,\quad {\rm at} \  68\%\  {\rm CL} \ ,\\
&& 18 <\fnl^{\rm loc.,\, BOSS}< 86\,, \quad {\rm or} \quad\fnl^{\rm loc.,\, BOSS}= 52 \pm 34 \ , \quad {\rm at} \  68\%\  {\rm CL} \ . \label{fnllocconst}
\eea
These constraints show no {significant} evidence of non-Gaussianity.
We perform a few additional checks on the viability of these results. The $\chi^2$ of the best fit of the model to the data, considering only the parameters that are relevant as degrees of freedom, has a sufficiently large corresponding $p$-value {$\sim 4\%$}.  Results for the joint analysis for $\fnl^{\rm equil.}-\fnl^{\rm orth.}$ give very similar results and the 68\% CL contour plot is given in Fig.~\ref{fig:contour}. Contour plots {for the posteriors} associated to these analyses are presented in~App.~\ref{sec:plots}.

\begin{figure}[t]
  \centering
  \begin{minipage}[t]{.33\textwidth}
  \vspace{0pt}
  \flushleft
  \includegraphics[width=1.3\textwidth]{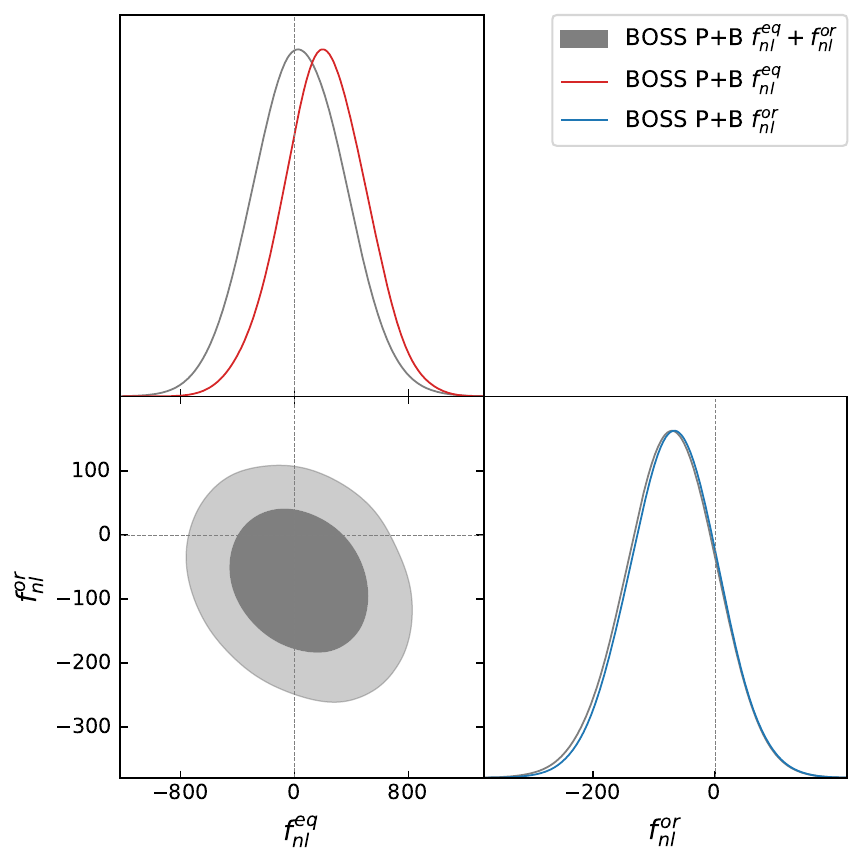}
  \end{minipage}%
  \begin{minipage}[t]{.66\linewidth}
  \vspace{40pt}
  \hspace{.9in}
  \normalsize
  \begin{tabular}{|c|c|c|c|} 
  \hline
  	&	 BOSS 	 	&  WMAP 	&  Planck  \\ \hline 
  $\fnl^{\rm equil.}$		&	{$207  \pm 292 $}  & $51 \pm 136 $ & $-26 \pm 47 $	  \\ 
  $\fnl^{\rm orth.}$	& $-68  \pm 73$ & $ -245 \pm 100$ & $ -38 \pm 24$			 \\
 $\fnl^{\rm loc.}$	&  $52 \pm 34$ & $37.2\pm 19.9$ & $-0.9\pm 5.1$	\\
  \hline
   \end{tabular}\\ 
  \end{minipage}
  
  \caption{\small  {\it Left:}   $\fnl^{\rm equil.}-\fnl^{\rm orth.}$ contour for the joint analysis (grey) and for the single $\fnl$ analyses (red) and (blue) respectively. {\it Right:} 68\%-confidence intervals for the BOSS analysis, as well as the WMAP~\cite{WMAP:2012fli} and Planck~\cite{Planck:2019kim} final results. We find no evidence of primordial non-Gaussianity.}
  \label{fig:contour}
  \end{figure}

It is interesting to investigate how it is possible to obtain such powerful limits. For the BOSS survey, we have an effective volume of about {$V_{\rm eff}\simeq 2.4 \,({\rm Gpc}/h)^3$}. The wavenumber $k_{\rm shot}$ at which the shot noise is comparable to the cosmic variance\footnote{{This estimate is done using the linear halo monopole power spectrum $P_{11,0}^h $ by computing $\bar n \, P_{11,0}^h ( k_{\rm shot}) = 1$ using the equations in \secref{btrssapp}, $b_1 = 2.2$, $\bar n=4 \cdot 10^{-4} \ (\hinvMpc)^3$, and $f = 0.78$ at the redshift $ z = 0.57$.}} is about {$k_{\rm shot} \simeq 0.33 \hinvMpc$} and therefore the number of available modes of the BOSS survey is $N_{\rm modes}\sim V_{\rm eff} \int^{k_{\rm shot}}d^3k/{(2\pi)^3}\simeq V_{\rm eff} k_{\rm shot}^3/(6\pi^2)\sim {1.5 \cdot 10^6}$.
The number of modes we use, since our {$\kmax\simeq 0.23 \hinvMpc$}, is thus $N_{\rm modes}\sim 0.5\cdot 10^6$.
We expect a bound of order $\fnl\zeta\lesssim 1/{\sqrt{N_{\rm modes}}}$: given that the amplitude of the primordial curvature fluctuation $\zeta$ is $\sim 3\cdot 10^{-5}$, we find an expected limit $\fnl\lesssim 50$. Of course, this estimate does not take into account the information absorbed into the EFT parameters, {the potential effect of shot noise which is not completely negligible at our wavenumbers}, {or of relative order one factors that go into the normalization of $\fnl$ for different shapes}, but still, it gives a sense of the {constraining} power of this LSS survey.

It is worthwhile to put our results in a broader context. Within LSS, the local shape has been analyzed in~\cite{Slosar:2008hx}, {and more recently in~\cite{Castorina:2019wmr,Mueller:2021tqa}}, and the local and the equilateral in~\cite{Xia:2011hj}. All references analyze only the power spectrum of several data sets, and use the non-local bias~\cite{Dalal:2007cu,Verde:2009hy,Schmidt:2010gw} that is induced by primordial non-Gaussianities, with a characteristic {and prominent} $k^{-2}$ scale-dependence for the local shape. Ref.~\cite{Xia:2011hj} obtains competitive constraints, but using information from the non-linear regime that depends on the astrophysics modeling. {Refs.~\cite{Castorina:2019wmr,Mueller:2021tqa} use an optimal estimator for non-Gaussianity and analyze the higher-redshift eBOSS quasar sample.} {Thus, it makes the most sense to compare our results} with~\cite{Slosar:2008hx}, which finds, on the DR5 LRG spectroscopic sample, $\fnl^{\rm loc.}=70^{+73}_{-84} \ {\rm at} \  68\%\  {\rm CL}$. When we analyze only the power spectrum {without the perturbativity prior (which is described in \secref{ppsec})} with the modeling provided by the EFTofLSS and on the {DR12 galaxy sample}, we find {$\fnl^{\rm loc.}= 57 \pm 57 \ {\rm at} \  68\%\  {\rm CL}$}. The two results look quite consistent, given that there are differences in the data version, modeling of the signal, and the choice of scales analyzed.
The addition of the bispectrum {together with the perturbativity prior} gives a remarkable error reduction of {$\simeq1.7$}. 

Perhaps even more interesting is the comparison with the bounds from the CMB. Planck constraints~\cite{Planck:2019kim} are still quite superior, but only by a factor that ranges from {$\simeq3.0$ for $\fnl^{\rm orth.}$, $\simeq 6.7 $ for $\fnl^{\rm loc.}$ to $\simeq6.2$ for $\fnl^{\rm equil.}$}.
In fact,  for the orthogonal shape, our results are already stronger by a factor of {$\simeq 1.4$} than the final results of WMAP~\cite{WMAP:2012fli}, {and} they are just {$\sim 41 \%$} weaker for the local shape and {$\sim 53\%$} for the equilateral shape.
Given that BOSS happened prior to the large program of upcoming LSS experiments, these results are of great hope for the future power of LSS, perhaps through specifically designed surveys, in constraining primordial non-Gaussianities.  {As an illustration, the outlook for the nearest upcoming survey, DESI~\cite{2013arXiv1308.0847L}, is quite promising, where we can expect reductions of the error bars by factors of $2.5$, $2.3$, and $5.2$, respectively for the equilateral, orthogonal, and local shapes \cite{Braganca:2023pcp}.\footnote{{Assuming control over systematics in the DESI emission line galaxy sample.}} } 

We also comment on the improvement from the addition of the one-loop prediction on top of the tree-level one for the bispectrum. We find that, taking $k_{\rm max} = 0.08 \ h / {\rm Mpc}$ as a reference for the tree-level analysis, our $f_{\rm NL}$ constraints obtained with the one-loop theory up to $k_{\rm max} = 0.23 \ h / {\rm Mpc}$ improve over the tree-level ones by a factor {$\sim 1.4$} on the equilateral shape and {$\sim 2.5$} on the orthogonal shape. {This is summarized in \tabref{noshotnoise}.}

 Finally, we comment on the effect of the shot noise in the BOSS sample on our results.  This is an interesting consideration since the shot noise is essentially determined by the number density of objects in the survey, and thus is (within reason) a part of the experimental design choice.  As mentioned earlier, the optimal limit on $f_{\rm NL}$ to expect from the BOSS survey at the $k_{\rm max}$ up to which we analyze the data is better by a factor of few than the limit we have obtained.  One reason for this {could be} that at large $k$, shot noise starts to swamp out the usable signal (it is approximately 50\% of the cosmic variance at $\kmax\simeq 0.23 \hinvMpc$).  {Another reason simply comes from the fact that we have to marginalize over the unknown EFT parameters.} In order to better understand this, in \tabref{noshotnoise} we show the size of the error bars on $f_{\rm NL}^{\rm equil.}$ and $f_{\rm NL}^{\rm orth.}$, using the tree-level and one-loop bispectrum, for the real BOSS data and for two synthetic BOSS experiments.  The first synthetic BOSS data are generated by our one-loop best-fit theory model, including the realistic mean number density of galaxies $\bar n=4 \cdot 10^{-4} \ (\hinvMpc)^3$.  This data is analyzed, using specifically the value $\bar n=4 \cdot 10^{-4} \ (\hinvMpc)^3$, with an analytic diagonal covariance matrix.  The second synthetic BOSS data are generated by our one-loop best-fit theory model, but with a value $\bar n=1 \ (\hinvMpc)^3$ such that the shot noise is effectively zero.  These data are analyzed, but now using the value $\bar n=1 \ (\hinvMpc)^3$, also with an analytic diagonal covariance matrix to simulate a hypothetical experiment with negligible shot noise.  
Additionally, for this case, we use the same priors as the previous cases but with the priors involving $\bar n$ rescaled appropriately.  
The synthetic BOSS data with realistic value of $\bar n$ was done as a consistency check, since it should give approximately the same results as the real BOSS analysis, which indeed is true.  {We see that overall, the error bars go down from the realistic analysis to the ideal analysis, by about {$34\%$} for $f_{\rm NL}^{\rm equil.}$ and {$40\%$} for $f_{\rm NL}^{\rm orth.}$ at one loop.  We also see that the \emph{relative} gain from using the one-loop bispectrum over the tree-level bispectrum improves (but not radically), suggesting that at the $k_{\rm max}$ that we use, we are not primarily dominated by shot noise.}


\begin{table}[t]
  \centering
  \normalsize
  \begin{tabular}{|c|c|c|c|c|c|c|}
  \hline
 &  \multicolumn{3}{ |c| }{$\sigma_{\rm equil.}$}  & \multicolumn{3}{ |c| }{$\sigma_{\rm orth.}$} \\ 
  \hline
  	&	1-loop bisp.&  tree bisp.  & \%	&  1-loop bisp.&  tree bisp.  & \%	\\ \hline 
  real BOSS		&	$292$ &  $420$  & $30$ & $73$	& $180$  & $59$   \\  \hline
  syn. BOSS, real $\bar n$	 & $299$ &  $420$ &  $29$ &  $ 93$		&   $168$ &   $ 45$	 \\
  syn. BOSS, large $\bar n$ 	&  $198$ &  $293$ &  $32$ &   $56$      &  $117$  &  $52$	 \\  \hline
   \end{tabular} 
  \caption{\small  {We give the improvements made when using the one-loop bispectrum (labeled `1-loop bisp.' in the table, $k_{\rm max} = 0.23 \, \hinvMpc $) compared with the tree-level bispectrum (labeled `tree bisp.' in the table, $k_{\rm max} = 0.08 \, \hinvMpc $) for the errors on $f_{\rm NL}^{\rm equil.}$ and $f_{\rm NL}^{\rm orth.}$ ($\sigma_{\rm equil.}$ and $\sigma_{\rm orth.}$ respectively) for various scenarios.  In the first line, we analyze the real BOSS data described in this paper (the one-loop bispectrum columns are the results given in this paper).  In the second line, we analyze synthetic (labeled `syn.' in the table) data generated by our theory model with cosmological parameters determined by our best fit, including the realistic shot noise.  Finally, in the third line, we analyze the same synthetic data, but with a large value of $\bar n$, such that the shot noise is effectively zero.    Columns marked \% give the percentage decrease in error bars due to using the one-loop bispectrum.    All cases use the one-loop power spectrum.  In general, we see that the EFTofLSS at one loop is expected to perform even better for experiments where the number density of objects is larger than that of BOSS. }  }
  \label{noshotnoise}
  \end{table}

{For reference, we show in Fig.~\ref{fig:bestfit} the best-fit plots of the power spectrum and bispectrum of BOSS galaxies with the EFTofLSS at one loop. 
To illustrate the sensitivity of the observables to primordial non-Gaussianity, we also show the departures from the best-fits within the $\sim 1\sigma$ precision on $f_{\rm NL}$ for the three shapes studied in this work. }

\begin{figure}\centering
\includegraphics[width=.56\columnwidth]{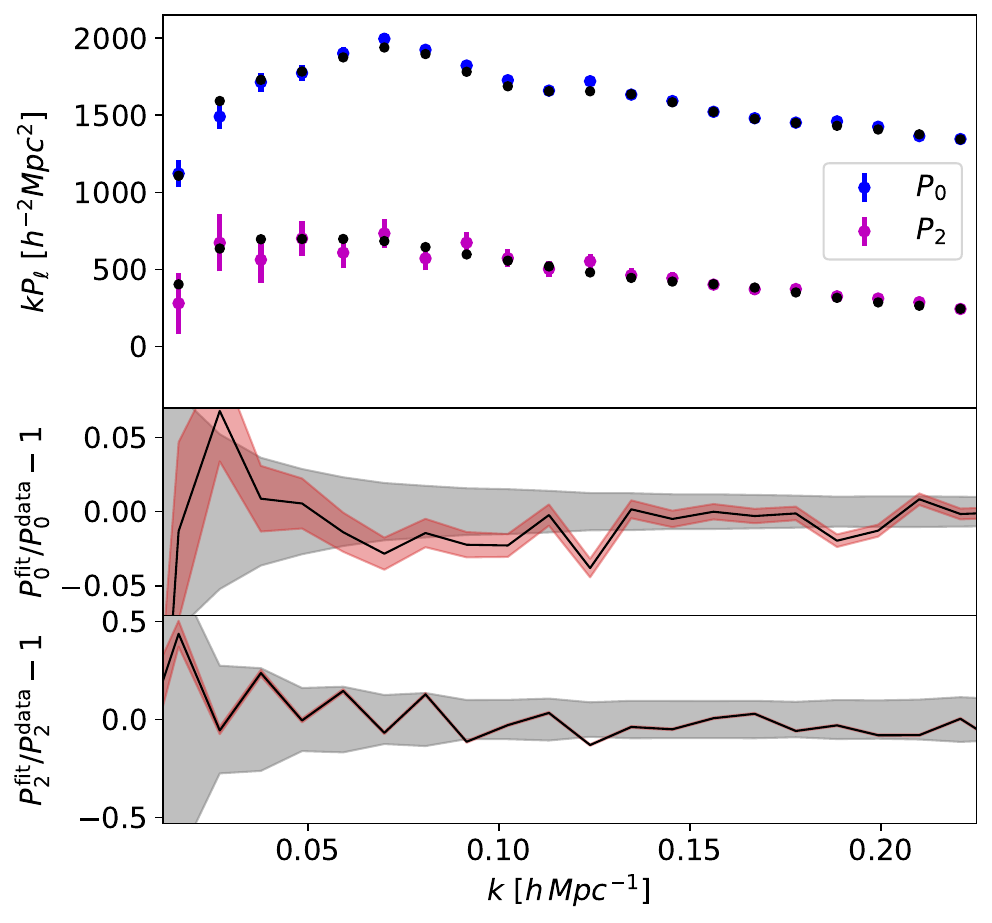}
\includegraphics[width=.99\columnwidth]{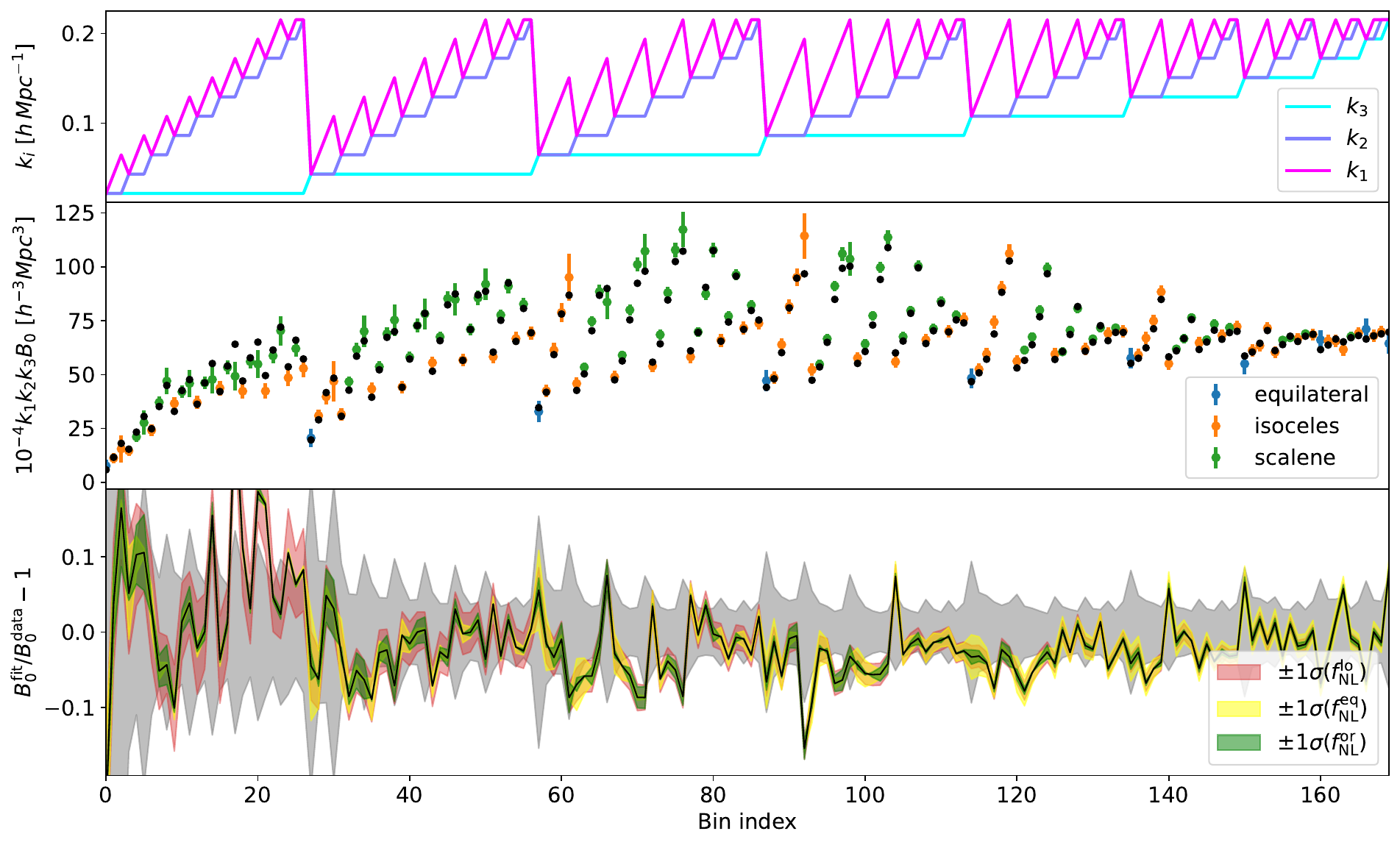}
\caption{\small {{Best fits (black dots) and residuals (black lines) of the power spectrum (upper plot) and bispectrum (lower plot) of BOSS galaxies with the EFTofLSS at one loop. 
The diagonal error bars of the data are shown in comparison (error bars around data points and gray regions in the residual plots). 
To avoid clutter, we only present CMASS NGC and only show the measurements with binning $12k_f$. 
In the lower plot for the bispectrum, the three sides of the central triangle in each bin are shown in the top panel. 
To illustrate the sensitivity of the galaxy statistics on $f_{\rm NL}$, we also show the regions comprising deviations from the best-fit within the $\sim 1\sigma$ precision obtained on $f_{\rm NL}$. }} }
\label{fig:bestfit}
\end{figure}


\paragraph{A note of warning:} We end this section of the main results with a final note of warning. It should be emphasized that in performing this analysis, as well as the preceding ones using the EFTofLSS by our group~\cite{DAmico:2019fhj,Colas:2019ret,DAmico:2020kxu,DAmico:2020ods,DAmico:2020tty,Zhang:2021yna}, we have assumed that the observational data are not affected by any unknown systematic error or undetected foregrounds. In other words, we have {used the publicly available BOSS catalogs.} Given the additional cosmological information that the theoretical modeling of the EFTofLSS allows us to exploit in BOSS data, it might be worthwhile to investigate if potential undetected systematic errors might affect our results. We leave an investigation of these  issues to future work.

\vspace{1cm}

\noindent {\bf Note added:} {While the initial arXiv preprint of this} paper was being finalized, {the by now published preprint of} \cite{Cabass:2022wjy} appeared a few days before this work. Ref.~\cite{Cabass:2022wjy} also analyzes the equilateral and orthogonal shape, but not the local shape {(although this study was posted to the arXiv a few months later by the same authors in the by now published \cite{Cabass:2022ymb})}. Regarding the overlapping shapes, the main difference is that \cite{Cabass:2022wjy} uses the tree-level prediction for the bispectrum, while we use the one-loop prediction. 
Because of this, our bounds are much stronger (considering that we are analyzing the same experiment). 
In order to compare, we analyzed the equilateral shape using only the tree-level prediction, finding overall agreement.  {The most optimistic numbers quoted in \cite{Cabass:2022wjy}, which assume {dark-matter halo relations that fix the quadratic biases in terms of $b_1$}, are $\fnl^{\rm equil.}= 260 \pm 300$ and $\fnl^{\rm orth.}= -23 \pm 120$.  If we also fix the quadratic bias, we obtain the even stronger constraints {$\fnl^{\rm equil.}= -119 \pm 233$}, {$\fnl^{\rm orth.}= -107 \pm 56$}, and {$\fnl^{\rm loc.}= 47 \pm 31$} (all numbers quoted here are at $68\%\  {\rm CL} $).  This again shows the constraining power of the one-loop bispectrum and the perturbativity prior.}  {However, we do not think that fixing quadratic biases is a justified procedure from the EFTofLSS point of view (for a study of the effect of futuristic but perhaps realistic priors on bias parameters, see \cite{Braganca:2023pcp}). Fixing biases completely would make the EFTofLSS no longer manifestly correct.}

\section*{Acknowledgements}

\noindent  We thank Babis Anastasiou, Diogo Bragan\c{c}a, Yaniv Donath, Henry Zheng for much help in parts related to this project, and J.J. Carrasco for conversations.  M.L. acknowledges the Northwestern University Amplitudes and Insight group, Department of Physics and Astronomy, and Weinberg College, and is also supported by the DOE under contract DE-SC0021485. Most of the measurements were performed with the help of the CINECA high performance computing resources.
Part of the analysis was performed on the HPC (High Performance Computing) facility of the University of Parma, whose support team we thank, part on the computer clusters LINDA $\&$ JUDY in the particle cosmology group at USTC, for which PZ is grateful to Yi-Fu Cai, Jie Jiang, and Dongdong Zhang for their support, and part on the Euler cluster of ETH Zurich, whose support team we thank.

%
%

%

\appendix

\section{Non-Gaussian bias parameters} \label{ngbiasapp}

 For the non-Gaussian bias parameters $b_1^{\fnl}$ and $b_2^{\fnl}$, we use the forms suggested in  \cite{Baldauf:2010vn}
\begin{align}
\begin{split} \label{eq:astro_relations}
w_\beta  b_1^{\fnl} & = 2    \delta_c ( b_1 -1)  \\ 
w_\beta  b_{2}^{\fnl} & = -  2 \left(  b_1 -1 - \frac{13}{21} \delta_c (   b_1 -1) - 2 b_5 \delta_c  \right) \ . 
\end{split}
\end{align}
where $\delta_c = 1.686$ is the critical collapse density.

\begin{figure}[t]
  \centering
   \includegraphics[width=0.7\textwidth]{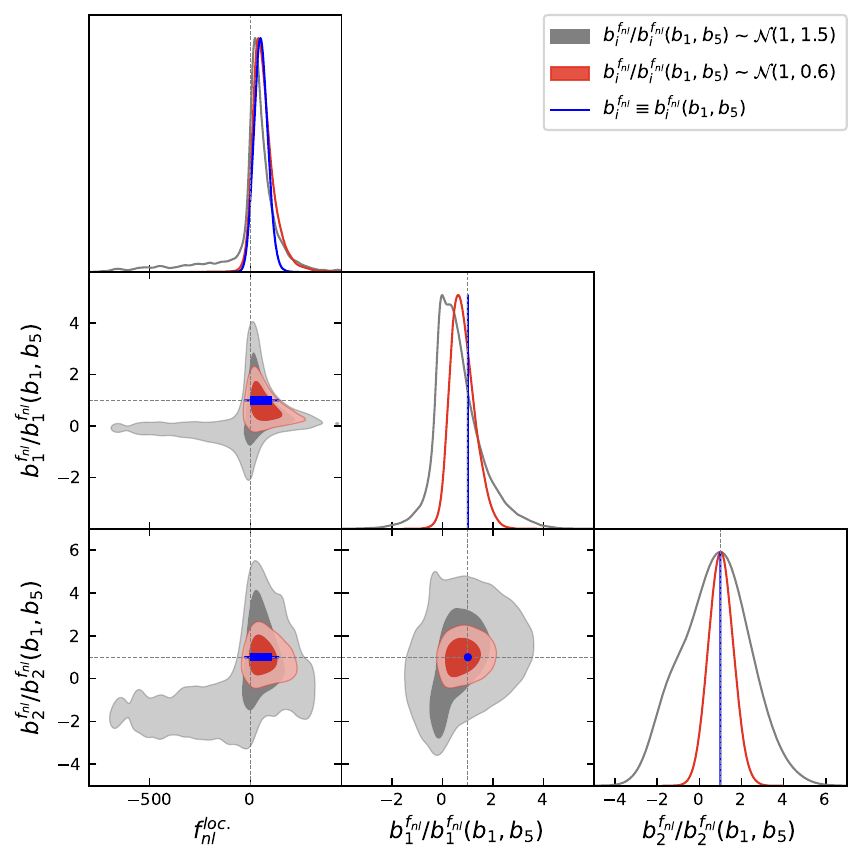}
  \caption{\small We plot $\fnl^{\rm loc.}$ posteriors from the joint power spectrum and bispectrum analysis of BOSS data for various choices of prior on the non-local biases $b_1^{\fnl}$ and $b_2^{\fnl}$: centering them on the universality relations~\eqn{eq:astro_relations}, we either impose a Gaussian prior of relative width $0\%$, $60\%$, or $150\%$.  {As described in the main text, the constraints on $\fnl^{\rm loc.}$ weaken as we widen the prior.  For the fixed bias relations, the thick blue line is the $1\sigma$ region, and the thinner blue line is the $2\sigma$ region.  }  $\mathcal{N} ( a , b )$ in the legend stands for a Gaussian distribution of mean $a$ and standard deviation $b$.  {Because the posterior for the 150\% prior is highly non-Gaussian, we show the plot with Gelman-Rubin $R-1 = 0.04$.}}
  \label{fig:freenonlocal}
  \end{figure}

Since, in our analysis, \eqn{eq:astro_relations} relates the value of the non-local biases to the value of $\fnl$, it is important for them to be accurate for an accurate limit on $\fnl$. However, it is probably too optimistic to assume that they are exact. We test the role of possible inaccuracies in the following way. For $\fnl^{\rm equil.}$ and $\fnl^{\rm orth.}$, we analyze the data by setting $b_1^{\fnl}$ and $b_2^{\fnl}$ to zero, and find that the results are practically unchanged. For $\fnl^{\rm loc.}$, we expect a larger effect from these terms, as the scale dependence of the biases is more prominent. If we let the non-local biases scan with a Gaussian prior centered around the value obtained in \eqn{eq:astro_relations} and with a width which is {60\%} of the same value to account for inaccuracies of the modeling, we find that the result on BOSS data is changed to {$\fnl^{\rm loc.} = 64^{+30}_{-60}  \,, {\rm at} \  68\%\  {\rm CL} $}, while with {the fixed bias relations in \eqn{eq:astro_relations}} we had obtained  {$\fnl^{\rm loc.}= 52 \pm 34 \,, {\rm at} \  68\%\  {\rm CL} $}.\footnote{{As just described, letting the non-local biases vary by $60\%$ when including the bispectrum leads to a {$\sim 32\%$} increase in the error bar of $\fnl$.
We can compare this to a power-spectrum-only analysis.  In this case, $b_1^{\fnl}$ is completely degenerate with $\fnl$, appearing only in the combination $b_1^{\fnl} \fnl$.  Because of this, we also put a flat prior $b_1^{\fnl} \fnl > 0.01$ because the limit $b_1^{\fnl} \fnl  \rightarrow 0$ corresponds to an infinite phase-space prior volume.
We find, in the analysis with only the power spectrum and letting the non-local biases vary by $60\%$, that the error bar on $\fnl$ increases by $\sim 60\%$ as expected. This illustrates that the bispectrum (through \eqn{k2kerexp}) breaks some degeneracies between $\fnl$ and the non-local biases.} } This shift is entirely dominated by the freedom in the linear non-local bias, but, as we see and considering the relatively large prior that we have allowed, it is not a large effect. 
{We therefore conclude that our results are robust against mild uncertainties in the relation described by \eqn{eq:astro_relations}.  }

 {In \figref{fig:freenonlocal}, we show the posteriors of $\fnl^{\rm loc.}$ obtained varying $b_1^{\fnl}$ and $b_2^{\fnl}$ with Gaussian priors centered on the values given by \eqn{eq:astro_relations} with relative widths of $60\%$ and $150\%$.  For large priors like $150\%$, we can see that the posterior becomes non-Gaussian, and takes on a characteristic hyperbola shape, due simply to the fact that we are actually bounding mostly the products $b_1^{\fnl} \fnl^{\rm loc.}$ and $b_2^{\fnl} \fnl^{\rm loc.}$.  The resulting constraints on $\fnl^{\rm loc.}$ weaken slightly: the $60\%$ number was quoted above, and for the prior of size $150\%$ we obtain {$\fnl^{\rm loc.}= 9^{+110}_{-31} $ at   $68\%\  {\rm CL} $}.  Notice that while the one-dimensional posterior for the $150\%$ prior on $\fnl^{\rm loc.}$ appears more narrow than the posterior with fixed bias relations, this is only because the former is non-Gaussian: indeed the $68 \% \ {\rm CL}$ for the $150\%$ prior is larger.  We do not consider larger priors on \eqn{eq:astro_relations} because of potential contamination from prior volume (or phase-space projection) effects \cite{Barreira:2022sey}.  
All in all, it seems to us that, for bias models up to 60\% or even 150\% different from \eqn{eq:astro_relations}, our results are robust against these kinds of uncertainties in \eqn{eq:astro_relations}.
For reference, quoting results for CMASS NGC, our current analysis gives {$b_1^{\fnl} \fnl^{\rm loc.} = 272 \pm 179$ and $b_2^{\fnl} \fnl^{\rm loc.} = 389 \pm 285$ at $68\%$ CL. }}

\section{Likelihood \label{sec:likelihood}}

\paragraph{Data:}
From the SDSS-III BOSS DR12 galaxy sample~\cite{BOSS:2016wmc}, we make use of the power spectrum and the bispectrum that we measure as follows. 
To each galaxy we assign the standard FKP weights for optimality together with the correction weights described in~\cite{Reid:2015gra} for BOSS data and in~\cite{Kitaura:2015uqa} for the patchy mocks. 
All celestial coordinates are converted to comoving distances assuming $\Omega_m^{\rm fid} = 0.310$. 
The power spectrum and the bispectrum are measured with the estimator described in~\cite{Feldman:1993ky,Yamamoto:2002bc,Yamamoto:2005dz,Bianchi:2015oia} and~\cite{Scoccimarro:1997st,Sefusatti:2005br,Baldauf:2014qfa,Scoccimarro:2015bla} respectively, using the code \code{Rustico}~\cite{Gil-Marin:2015sqa},\footnote{\href{https://github.com/hectorgil/Rustico}{https://github.com/hectorgil/Rustico}}  {and we note that a Poisson shot noise is subtracted directly at the level of the estimator.}
For all the sky cuts we use a box consisting of $512^3$ cells of side length $L_{\rm box} = 3500 \Mpcinvh$, with Piecewise Cubic Spline particle assignment scheme and grid interlacing~\cite{Sefusatti:2015aex}. 
The measurements are binned in $\Delta n = 6$ units of the fundamental frequency of the box $k_f$, starting from the bin centered at $n_{\rm min} = 9$, up to the one centered on $n_{\rm max} = 123$, which correspond in frequencies to bins of size $\Delta k \simeq 0.0108 \hinvMpc$, with first and last bins centered on $k_{\rm min} = 0.016 \hinvMpc$ and {$k_{\rm max} \simeq 0.221  \hinvMpc$}, respectively. 
Importantly, we keep all bins {whose centers form a closed triangle}}.
Explicitly, we select the bins whose centers satisfy:
\begin{equation}\label{eq:tcenters}
  \begin{split}
   (n_1, n_2, n_3)\,, \quad & n_1, n_2, n_3 = n_{\rm min}, n_{\rm min}+\Delta n,  \dots , n_{\rm max}\, , \\
  & \text{if } n_1 \leq n_2 \leq n_3 \text{ and }  n_3 \leq n_1 + n_2 \, .
  \end{split} 
\end{equation}
In this analysis, we use two redshift cuts $0.2<z<0.43$ and $0.43 < z < 0.7$, namely for LOWZ and CMASS, respectively, with two galactic cuts NGC and SGC each, for a total of four skies.

For the local shape, we find that for N-series, we can reliably take $k_{\rm min}=0.005 \hinvMpc$ for the power spectrum. However, for the data, as there can be effects at large scales like {observational systematics} beyond the ones simulated in the N-series (see for instance~\cite{Noorikuhani:2022bwc}), we use $k_{\rm min}=0.01 \hinvMpc$ throughout. 
For the orthogonal shape and when fitting jointly with the equilateral shape, we re-bin the data and the covariance such that $\Delta n = 12$.

\paragraph{Likelihood:}
We use the following likelihood $\mathcal{L}$ to describe the data:
\be
-2 \log( \mathcal{L} ) = (D-T) \cdot C^{-1} \cdot (D-T) \, .
\ee
Here $D$ is the data vector, constructed from the measurements of the power spectrum monopole and quadrupole, concatenated with the bispectrum monopole. 
$T$ is the corresponding EFTofLSS prediction, as described in \appref{sec:theory}, containing also additional modeling aspects presented below in this appendix.
Finally, $C^{-1}$ is the inverse covariance built from the 2048 patchy mocks, where we first concatenated the power spectrum multipoles with the bispectrum monopole of each realization. 
Given the finite number of mocks used to estimate the inverse covariance, we correct it with the Hartlap factor~\cite{Hartlap:2006kj}. 
We have checked that, although the Hartlap factor is not a small correction, our estimation of the inverse covariance is unbiased: instead of the analysis with $40 + 1015$ bins (per skycut/patchy mock suite), we have explicitly checked that we obtain similar results by restricting the analysis to fewer bins ($100, 200, \dots$) constructed as the linear combinations that maximize the signal-to-noise of $f_{\rm NL}$, following the method outlined in~\cite{Philcox:2020zyp}.

\paragraph{Posterior sampling:}
To sample the posteriors, we use the partially-marginalized likelihood as described in~\cite{DAmico:2019fhj,DAmico:2020kxu}, {which already includes the bispectrum, as it was analyzed first in~\cite{DAmico:2019fhj}}.
In this likelihood, the EFT parameters appearing only linearly in the predictions are marginalized analytically. 
We fix all cosmological parameters, but $f_{\rm NL}$, to Planck preferred values~\cite{Planck:2018vyg}. 
We are thus left to sample using MCMC only $f_{\rm NL}$, and, for each skycut, $b_1$, $b_2$, $b_5$. Therefore the sampling of the likelihood is extremely fast: a chain typically runs in~$10$ minutes on a single-core processor. 
All our results are presented with Gelman-Rubin convergence $R-1 < 1\%$, and are obtained with the Metropolis-Hasting algorithm as implemented in MontePython v3.3. 
{We have checked that our results are robust upon change of the sampler to PolyChord~\cite{Handley:2015fda,Handley:2015vkr}. }

\begin{table}[] \centering
\begin{tabular}{|c|c|}\hline
Parameter & Prior               \\ \hline
$b_1$     & Lognormal(0.8, 0.8) \\
$b_i$  ,  $ i = 2 , \dots , 11$    & $\mathcal{N}(0, 2)$ \\
$c_1^h$, $c_1^\pi$, $c_1^{\pi v}$	& $\mathcal{N}(0, 4)$ \\
$c^{\pi v}_3$ , $c^h_2$ , $c^h_3$, $c^h_4$, $c^h_5$, $c^{\pi}_5$, $c^{\pi v}_2$, $c^{\pi v}_4$, $c^{\pi v}_5$	& $\mathcal{N}(0, 2)$ \\
$c^{\rm St}_2$	& $\mathcal{N}(0, 4)$ \\
$(2/3) f c^{\rm St}_3$, $ c^{\rm St}_4$, $c^{\rm St}_5$,  $2 c^{\rm St}_9$, $- c^{\rm St}_7 - c^{\rm St}_9$, $ - c^{\rm St}_8 - 2 c^{\rm St}_9$, $2 c^{\rm St}_{12}$, $c^{\rm St}_{11}$, $c^{\rm St}_{10}$, $c^{\rm St}_{13}$    &  $\mathcal{N}(0, 2)$   \\
$c^{\rm St}_1$	& $\mathcal{N}(0, 0.2)$ \\
$c^{(222)}_1$	& $\mathcal{N}(0, 0.4)$ \\
$2c^{\rm St}_6$	& $\mathcal{N}(0, 1)$ \\
\hline
\end{tabular}
\caption{\label{tab:prior} \small {Priors on EFT parameters used in our analyses.  Here, we use the notation of \cite{DAmico:2022ukl}.  The various factors and combinations of parameters that appear above are the ones that appear in front of independent functional forms of the counterterm expressions.  See \cite{DAmico:2022ukl} for an explicit explanation.
Except for the linear bias $b_1$ for which we impose a lognormal prior centered on 0.8 with variance 0.8, we impose on nonlinear galaxy biases $b_i$, $i = 2, \dots , 11$, a normal Gaussian prior $\mathcal{N}$ centred on 0 with standard deviation 2. 
For the explicit scales appearing in counterterms, we use $k_{\rm M} = 0.7 \, \hinvMpc$, $k_{\rm R} = 0.25 \, \hinvMpc$, and $\bar n = 4 \cdot 10^{-4} \, (\Mpcinvh)^{3}$. 
For the EFT parameters $c_1^{\rm St}$, $c_6^{\rm St}$, and $c_{1}^{(222)}$, related to the $\mathcal{O}(k^0)$ stochastic terms, their priors are tighter assuming a distribution close to Poisson (while having explicitly set the $\mathcal{O}(k^0)$-limit of the loops to 0). 
We further impose a correlation on EFT parameters amongst the four skies of BOSS as explained~\cite{DAmico:2022osl}. }
}
\end{table}

\paragraph{Prior:}
We impose {an uninformative large flat prior on $f_{\rm NL}$.  {The priors on the EFT parameters used in this work are summarized in \tabref{tab:prior}. 
These priors are chosen such that nonlinear contributions in the EFTofLSS are varying within physical range provided order-of-magnitude estimates on their natural size. }  For $b_1$, which is positive-definite, we choose a lognormal prior, given by
\be
\mathcal{P} = \frac{1}{b_1 \sigma \sqrt{2 \pi}} e^{-\frac{(- \mu+ \log b_1 )^2}{2 \sigma^2}}
\ee
with $\mu = 0.8$ and $\sigma^2 = 0.8$.  The median value of $b_1$ is 2.23, and the 68\% confidence interval centered on that is [0.91, 5.4]. }
For the remaining EFT parameters,  {including $b_2$ and $b_5$}, we use Gaussian priors centered on $0$, with various widths $\sim \mathcal{O}(b_1)$, to keep them within physical range.\footnote{Note that $b_5$ is equal to $b_4$ in the notations of~\cite{DAmico:2019fhj}.}
{For the other parameters entering in the power spectrum, we use the same prior widths as described in~\cite{DAmico:2019fhj}, while for the additional parameters entering only in the bispectrum, we use prior widths as described in~\cite{DAmico:2022osl}.  {The only changes with respect to those references are for the stochastic parameters, where we use priors of widths $0.2$ on $c_{1}^{\rm St}$,  $1$ on $2 c_{6}^{\rm St}$, and $0.4$ on $c_1^{(222)}$}.}  This choice is made to account for contributions degenerate with shot noise, such as corrections from fiber collisions \cite{Hahn:2016kiy}.}  We also assume a small correlation between the skycuts given the small redshift evolution of the EFT parameters and differences between the selection functions, as described in~\cite{DAmico:2022osl}.  {This prior is complemented by an expectation imposed directly on the size of the loop, as explained in the main text. }  {For reference, we obtain the following constraints for our analysis with the one-loop bispectrum, but with no perturbativity prior: {$\fnl^{\rm equil.}= 4 \pm 307$, $\fnl^{\rm orth.}= -75 \pm 86$, and $\fnl^{\rm loc.}= 86 \pm 51$  at $ 68\%\  {\rm CL}$. }}

\paragraph{Modeling aspects:}
To make contact with observations, we extend the EFTofLSS predictions to additional modeling aspects. 
For the power spectrum, we correct for the Alcock-Paczynski effect~\cite{Alcock:1979mp}, window functions, and binning as done in \code{PyBird}~\cite{DAmico:2020kxu}, {and we also include the integral constraints as described in~\cite{deMattia:2019vdg}.\footnote{See \href{https://github.com/pierrexyz/fkpwin}{https://github.com/pierrexyz/fkpwin} for the practical implementation of the geometrical effects.  } }
For the bispectrum, the corresponding corrections are described in details in~\cite{DAmico:2022osl}. 
Here we give a quick summary of the additional modeling in the bispectrum. 
First, we implement the IR-resummation as in~\cite{DAmico:2022osl}. Second, the Alcock-Paczynski effect is corrected at tree level, {while we estimate it to be negligible in the loop}. 

Third, we implement the window function only at tree level and as approximately as in~\cite{DAmico:2019fhj}, and we estimate that the error, for the scales analyzed, is negligible with respect to BOSS error bars.  {Specifically, in \cite{DAmico:2019fhj, DAmico:2022osl}, it was shown that for the $\Lambda$CDM analysis, the error in the window function for the bispectrum monopole is negligible on N-series and BOSS.  To look at the effect specifically on $\fnl^{\rm loc.}$ (where one might worry that the effect is larger), we performed the following tests.  On N-series, we find $\fnl^{\rm loc.} = -4 \pm 4$ when setting $k_{\rm min} = 0.04 \hinvMpc$ for the bispectrum, and  $\fnl^{\rm loc.} = - 1 \pm 4$ with $k_{\rm min} = 0.01 \hinvMpc$ for the bispectrum, but with completely removing the window function from the bispectrum (for quick reference, our result in \eqn{simsresultslo} with the bispectrum $k_{\rm min} = 0.01 \hinvMpc$ and the approximate window treatment, we obtained $\fnl^{\rm loc.} = 3 \pm 4$).  On BOSS data, we find $\fnl^{\rm loc.} = 41 \pm 38$ when setting $k_{\rm min} = 0.04 \hinvMpc$ for the bispectrum (for quick reference, our result in \eqn{fnllocconst} with the bispectrum $k_{\rm min} = 0.01 \hinvMpc$ and the approximate window treatment, we obtained $\fnl^{\rm loc.} = 52 \pm 34$).  Comparing with our results \eqn{simsresultslo} and \eqn{fnllocconst}, we see that all of these shifts are less than $1/3$ of the BOSS error bars on $\fnl^{\rm loc.}$, which confirms that we are making a marginal error using the approximate window function.  For completeness, on the BOSS data, we also find $\fnl^{\rm loc.} = 4 \pm 33$ with $k_{\rm min} = 0.01 \hinvMpc$ for the bispectrum, but with removing the window function from the bispectrum.  This is a larger shift, but it is expected because completely removing the window function, rather than approximately implementing it, is expected to be a large error.  }

Finally, one can see from Eq.~\eqref{eq:tcenters} that several bins we analyze contain fundamental triangles that are not closed (i.e. the bin centers do not form a triangle, but there are closed triangles within the bins). 
To properly account for them, and also for the fully closed lower $k$-bins, a binning scheme for the theory model is needed. 
The procedure we implement is outlined in detail in~\cite{DAmico:2022osl}, and is found to be robust for the analysis of BOSS data. 
In brief, the tree-level part is binned exactly while the loop, being smaller, is evaluated on effective wavenumbers. 
In this work, we also bin exactly the pieces proportional to $f_{\rm NL}$, which turns out to be important especially for the local and the orthogonal shapes.

The goodness of all these approximations is confirmed by the fact that we find no  evidence of {significant} theoretical systematic error in the N-series `cutsky'   simulations, that are mimicking the geometry of BOSS CMASS NGC. As we analyze them using predictions for the bispectrum with approximate window function and IR-resummation, with our binning prescription and with the correction for Alcock-Paczynski effect only at tree level in the bispectrum, we conclude that our cosmological results are robust against the modeling aspects discussed here.

\section{Plots \label{sec:plots}}
\begin{figure}
\centering
\includegraphics[width=0.8\textwidth]{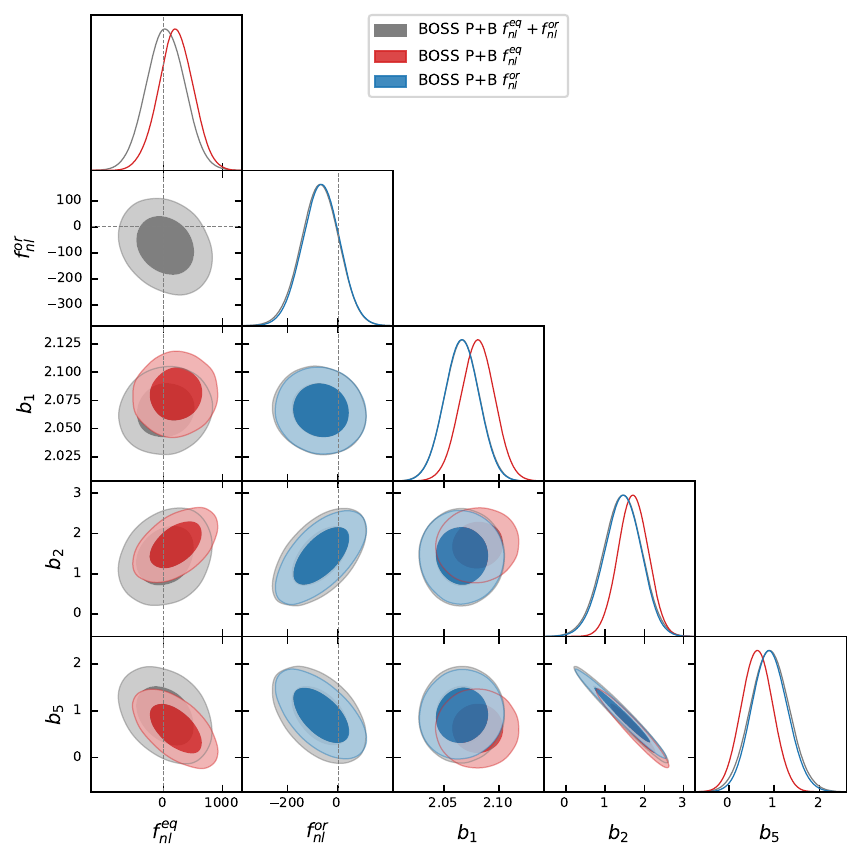}
\caption{\small  Triangle plots obtained fitting the BOSS data in an analysis where we vary, together with the EFT parameters, only  $\fnl^{\rm equil.}$ (red), only $\fnl^{\rm orth.}$ (blue), and jointly $\fnl^{\rm equil.}$ and $\fnl^{\rm orth.}$ (grey). The analysis on the BOSS data reveals no evidence of primordial non-Gaussianities of these kinds. $\fnl^{\rm equil.}$ and $\fnl^{\rm orth.}$ appear to have a correlation of {$\sim-0.23$}.}
\label{fig:equilortho}
\end{figure}

\begin{figure}
\centering
\includegraphics[width=.49\textwidth]{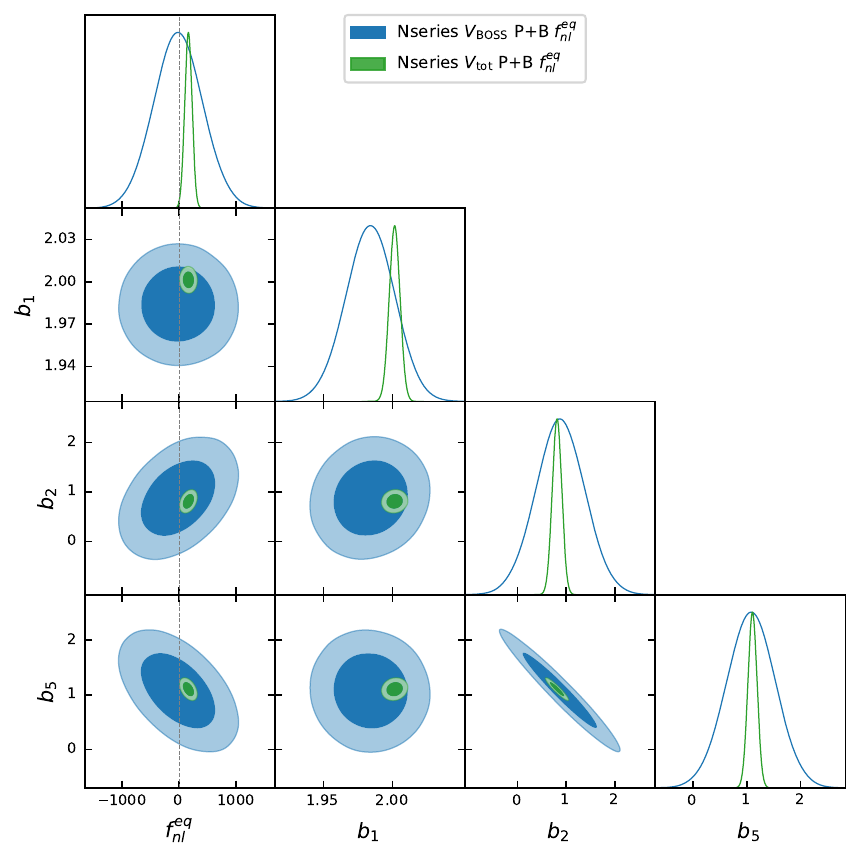}
\includegraphics[width=.49\textwidth]{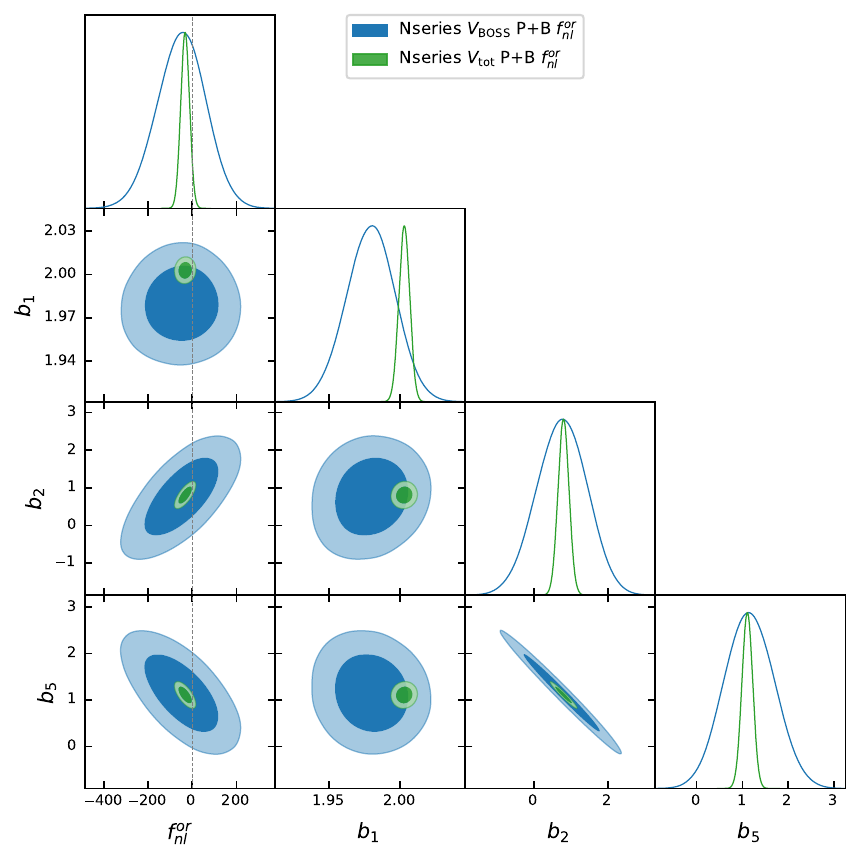}
\caption{\small   Triangle plots obtained fitting the N-series data ({with `$P$' being `power spectrum' and `$B$' being `bispectrum'}) in an analysis where we vary, together with the EFT parameters, $\fnl^{\rm eq.}$, or $\fnl^{\rm orth.}$. The N-series data represent the average over 84 boxes each of volume comparable to the BOSS one, {and we show the results obtained with the total covariance of volume $V_{\rm tot}$ or of volume corresponding to one box $V_{\rm BOSS}$. } This allows us to conclude that the analysis has no large {theoretical} systematic error. }
\label{fig:nseries_eq}
\end{figure}

\begin{figure}
\centering
\includegraphics[width=.8\textwidth]{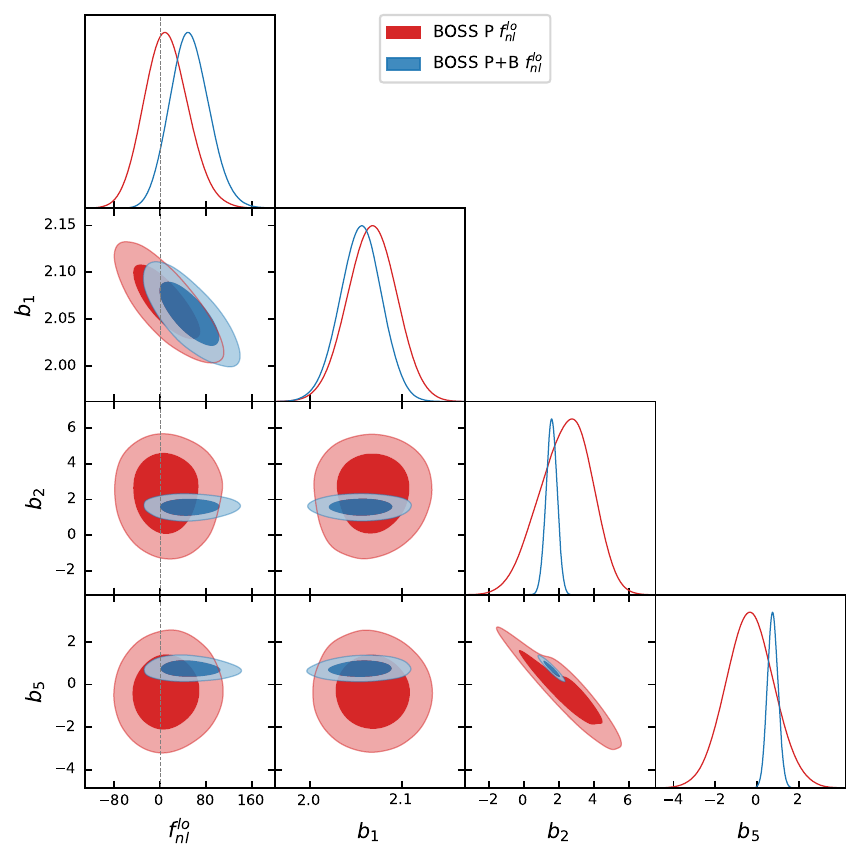}
\caption{\small   Triangle plots obtained fitting the BOSS data ({with `$P$' being `power spectrum' and `$B$' being `bispectrum'}) in an analysis where we vary, together with the EFT parameters, $\fnl^{\rm loc.}$.  The analysis on the BOSS data reveals no evidence of primordial non-Gaussianity of this kind.
}
\label{fig:local}
\end{figure}

\begin{figure}
\centering
\includegraphics[width=.8\textwidth]{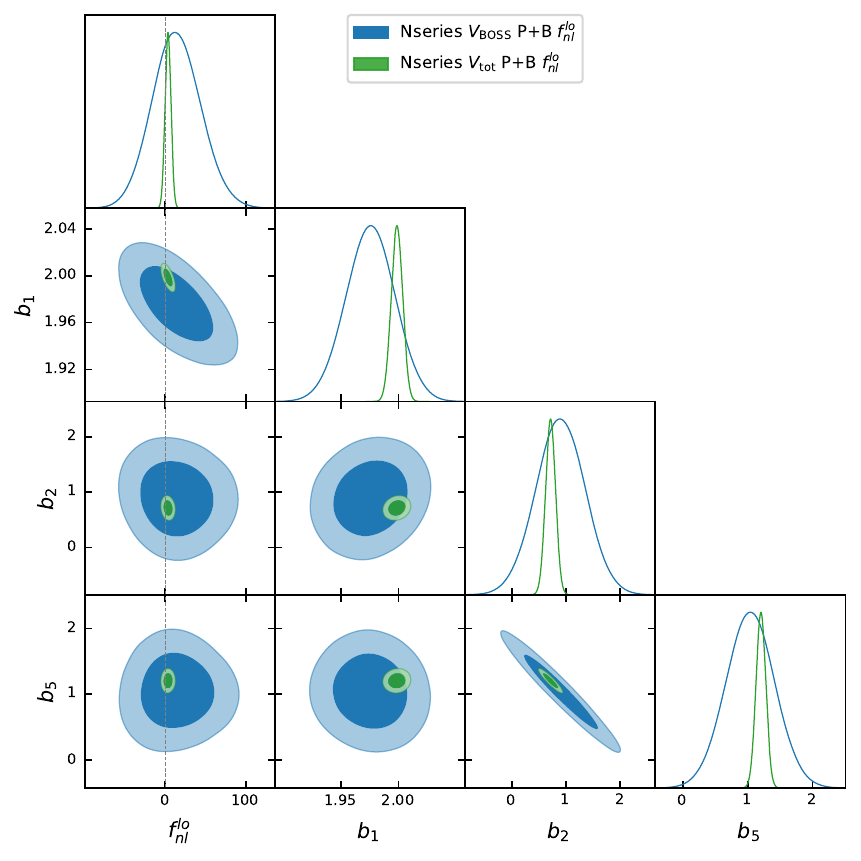}
\caption{\small   Triangle plots obtained fitting the N-series data ({with `$P$' being `power spectrum' and `$B$' being `bispectrum'}) in an analysis where we vary, together with the EFT parameters, $\fnl^{\rm loc.}$. The N-series data represent the average over 84 boxes each of volume comparable to the BOSS one, {and we show the results obtained with the total covariance of volume $V_{\rm tot}$ or of volume corresponding to one box $V_{\rm BOSS}$. } This allows us to conclude that the analysis has no large {theoretical} systematic error.
}
\label{fig:nseries_local}
\end{figure}

In Figs.~\ref{fig:equilortho},~\ref{fig:nseries_eq},~\ref{fig:local}, and~\ref{fig:nseries_local}, we present the posteriors obtained in the main analyses of this work. 
The shown EFT parameters are the ones of CMASS NGC. As the contours for the other smaller skycuts look similar except that they have larger error bars, we do not shown them for clarity.

\bibliographystyle{JHEP}
\small
\bibliography{references}

\end{document}